\documentclass[preprint]{aastex}
\usepackage{natbib}
\usepackage{amssymb}
\usepackage{lscape}
\usepackage{longtable}
\usepackage{url}
\usepackage{epsfig}

\shorttitle{``Merger-nova" emission from long-lived magnetar}
\shortauthors{Yuan et al.}
\slugcomment{}

\begin{document}
\title{Detectability of ``Merger-nova" emission from a long-lived magnetar in short gamma-ray bursts}
\author{Yong Yuan\altaffilmark{1}, Hou-Jun L\"{u}\altaffilmark{1}, Hao-Yu Yuan\altaffilmark{1}, Shuai-Bing
Ma\altaffilmark{2}, Wei-Hua Lei\altaffilmark{2}, and En-Wei Liang\altaffilmark{1}}

\altaffiltext{1}{Guangxi Key Laboratory for Relativistic Astrophysics, School of Physical Science
and Technology, Guangxi University, Nanning 530004, China; lhj@gxu.edu.edu} \altaffiltext{2}{School
of Physics, Huazhong University of Science and Technology, Wuhan 430074, China}

\begin{abstract}	
One possible progenitor of short gamma-ray bursts (GRBs) is thought to be from a double neutron
star (NS) merger, and the remnant of such a merger may be a supramassive NS, which is supported by
rigid rotation and through its survival of hundreds of seconds before collapsing into a black hole
(BH). If this is the case, an optical/infrared transient (namely merger-nova) is generated from the
ejected materials and it is powered by radioactive decay from $r$-process, spin-down energy from a
supramassive NS, as well as the magnetic wind from a newborn BH. In this paper, we systematically
search for the signature of a supramassive NS central engine by analyzing the X-ray emission of
short GRBs with internal plateau observed by {\em Swift}, and we find that five candidates of short
GRBs have such feature with redshift measurement. Then, we calculate the possible merger-nova
emission from those candidates for given the typical model parameters by considering the above
three energy sources, and compare its brightness with the sensitivity of some optical telescopes.
We find that the merger-nova emission of GRB 060801 in K-, r-, and U-bands with variations of
$M_{\rm ej}$ ($10^{-4}-10^{-2} M_{\odot}$), $\kappa$ ($0.1-10 ~\rm cm^{2}~g^{-1}$), and $\beta$
($0.1-0.3$) is very difficult to detect using the Vera C. Rubin, Panoramic Survey Telescope and
Rapid Response System (Pan-STARRS), the Zwicky Transient Facility, and the Roman Space Telescope
(Roman), except for the case of large ejecta mass $M_{\rm ej}=10^{-2} M_{\odot}$. However, we are
very hopeful to detect the merger-nova emission of GRBs 090515, 100625A, and 101219A using more
sensitive instruments, such as Vera C. Rubin, Pan-STARRS, and Roman. Moreover, the merger-nova
emission of GRB 160821B is bright enough to detect in our calculations, and it is also consistent
with current real observations of merger-nova emission.
\end{abstract}


\section{Introduction}
The merger of two neutron stars (NSs) are thought to be potential sources of producing
gravitational-wave (GW) radiation and short gamma-ray bursts (GRBs; Berger 2014 for a review). The
first ``smoking gun" evidence to support this hypothesis is the direct detection of GW170817 and
its electromagnetic counterpart (e.g., GRB 170817A and AT2017agfo) originating from the merger of a
binary NS system that was achieved via the collaboration of Advanced LIGO and Advanced VIRGO, as
well as space- and ground-based telescopes (Abbott et al. 2017a; Covino et al. 2017; Goldstein et
al. 2017; Kasen et al. 2017; Savchenko et al. 2017; Zhang et al. 2018). Binary NS systems play an
especially interesting role in the field of astrophysics, such as studying the equation of state
(EoS) of NSs, the origin of heavy elements produced through $r$-process nucleosynthesis, testing of
the equivalence principle, and constraining Lorentz invariance (Abbott et al. 2017b; Burns 2020;
Foucart 2020, for a review). However, the post-merge product of NS merger remains an open question,
and it is dependent on the total mass of the post-merger system and the poorly known EoS of NS
(Lasky et al. 2014; L{\"u} et al. 2015; Li et al. 2016).

From the theoretical point of view, binary NS mergers may form a black hole (BH), or a supramassive
NS that is supported by rigid rotation and through its survival for hundreds of seconds before
collapsing into a BH if the EoS of NS is stiff enough (Rosswog et al. 2000; Dai et al. 2006; Fan \&
Xu 2006; Metzger et al. 2010; Yu et al. 2013; Zhang 2013; Lasky et al. 2014; L\"{u} et al. 2015;
Gao et al. 2016), or even a stable Ns (Dai \& Lu 1998a,b). No matter what the remnant is, electro
magnetic (EM) transients are probably produced during the coalescence process, such as the short
GRB and its afterglow emission within a small opening angle (Rezzolla et al. 2011; Troja et al.
2016; Jin et al. 2018), and an optical/infrared transient generated from the ejected materials and
powered by radioactive decay from $r$-process with near-isotropic (Li \& Paczynski 1998; Metzger et
al. 2010; Berger 2011; Rezzolla et al. 2011; Hotokezaka et al. 2013; Rosswog et al. 2013).
Moreover, such a transient is also suggested to be referred to as "macronova" (Kulkarni 2005),
"kilonova" (Metzger et al. 2010), or "merger-nova" (Yu et al. 2013; Metzger \& Piro 2014).

In terms of observations, the "internal plateau" emission is defined as a nearly fairly constant
emission followed by an extremely steep decay in the X-ray light curve of GRBs (Troja et al. 2007).
Moreover, the X-ray and optical data during the plateau phase are required to be not simultaneously
consistent with the forward shock model when extracting the spectrum. The first smoking gun of such
an internal plateau was discovered by Troja et al. (2007) in long GRB 070110. Whereafter, a good
fraction of short GRBs with X-ray emission present a internal plateau emission observed by {\em
Swift}/XRT (Rowlinson et al. 2010, 2013; L\"{u} et al. 2015, 2017; Troja et al. 2019). This feature
at least strongly supports a unstable supramassive NS as the center engine of some short GRBs
before it collapses into BH (Dai et al. 2006; Gao \& Fan 2006; Metzger et al. 2008; Fan et al.
2013; Zhang 2013; Ravi \& Lasky 2014; L\"{u} et al. 2015; Gao et al. 2016; Chen et al. 2017; L\"{u}
et al. 2020; Sarin et al. 2020a,b). If this is the case, the main power of the transients from a
double NS merger should no longer be limited to $r$-process, and the spin energy of the magnetar
can be comparable to or even larger than the radioactive decay energy (Yu et al. 2013; Zhang 2013;
Metzger \& Piro 2014; Gao et al. 2017; Murase et al. 2018). Furthermore, the magnetic wind from the
accretion disk of a newborn BH would heat up the neutron-rich merger ejecta and provide additional
energy to the transients (Jin et al. 2015; Yang et al. 2015; Ma et al. 2018). Back to previous
studies, this additional energy source from the post-merger central engine had been adopted to
power a bright X-ray afterglow emission (Zhang 2013), or a bright optical/infrared emission (Yu et
al. 2013; Metzger \& Piro 2014; Gao et al. 2015; Jin et al. 2015; Ma et al. 2018). Although, there
are a number of papers that discuss the rate of kilonova detections in optical surveys, but they do
not consider the contributions of all possible energy sources (Scolnic et al. 2018; Andreoni et al.
2020; Sagu{\'e}s Carracedo, et al. 2020; Thakur et al. 2020).

One basic question is how bright the merger-nova is if we simultaneously consider the above three
lines of energy sources, i.e., r-process-powered, spin energy from a supramassive NS, and magnetic
wind from the accretion disk of a newborn BH. Whether this merger-nova can be detected by optical
telescopes in the future, such as the Vera C. Rubin Observatory's Legacy Survey of Space and Time
(Vera C. Rubin; Ivezi{\'c} et al. 2019), the Zwicky Transient Facility (ZTF; Bellm et al. 2019;
Graham et al. 2019), the Panoramic Survey Telescope and Rapid Response System (Pan-STARRS; Morgan
et al. 2012; Chambers et al. 2016), and Roman Space Telescope (Roman; Pierel et al. 2020).
Following the above consideration, we systematically search for the X-ray emission of short GRBs
with internal plateau. This feature is believed to be corresponding to the process of a short
lifetime supramassive NS as central engine then collapsing into BH. Then, we calculate the
merger-nova emission by considering the above three energy sources, and discuss the detectability
of merger-nova to compare with the detected limit of optical telescopes in the future. Throughout
the paper, a concordance cosmology with parameters $H_0=71~\rm km~s^{-1}~Mpc^{-1}$,
$\Omega_M=0.30$, and $\Omega_{\Lambda}=0.70$ is adopted.

\section{Basic model of merger-nova energy sources}
If the central engine of some short GRBs is a magnetar, for given an initial spin period $P_{\rm
i}$, its total rotational energy can be written as $E_{\rm rot}\sim 2\times
10^{52}I_{45}M_{1.4}R^{2}_{6}P_{\rm i,-3}^{-2}\rm erg$, where $M$, $R$, and $I$ are the mass,
radius, and moment of inertia of a neutron star, respectively. Throughout the paper, the convention
$Q_x = Q / 10^x$ is adopted in cgs units. The spin-down luminosity of the millisecond magnetar as a
function of time may be expressed by the magnetic dipole radiation formula equation (Zhang \&
M\'esz\'aros 2001; L\"{u} et al. 2018)
\begin{eqnarray}
L_{\rm sd} = L_{0}\left(1+\frac{t}{\tau}\right)^{\alpha}
\label{Lsd}
\end{eqnarray}
where $L_0 \simeq 1.0 \times 10^{47}~{\rm erg~s^{-1}} (B_{p,14}^2 P_{i,-3}^{-4} R_6^6)$ and $\tau
\simeq2 \times 10^5~{\rm s}~ (I_{45} B_{p,14}^{-2} P_{0,-3}^2 R_6^{-6})$ are the characteristic
spin-down luminosity and timescale, respectively. $\alpha$ is usually constant and equal to -2 for
a stable magnetar central engine when the energy loss of the magnetar is dominated by magnetic
dipole radiation (Zhang \& M{\'e}sz{\'a}ros 2001). However, $\alpha$ may be less than -2 if the
central engine is a supramassive NS whether the energy loss is dominated by magnetic dipole
radiation or a gravitational wave (Rowlinson et al. 2010; Lasky \& Glampedakis 2016; L\"{u} et al.
2018). The near-isotropic magnetar wind interacts with ejecta and is quickly decelerated. On the
other hand, the injected wind continuously pushes from behind and accelerates the ejecta. In our
calculation, we assume that the magnetar wind is magnetized, and the wind energy could be deposited
into the ejecta either via direct energy injection by a Poynting flux (Bucciantini et al. 2012), or
via forced reconnection (Zhang 2013). More details about the dynamical evolution of the ejecta will
be discussed in below.

The supramassive NS will collapse into a BH after a few hundred seconds when the degeneracy
pressure of the neutron cannot support the gravitational force due to its rotation energy loss. The
strongly magnetic field from the newborn BH will be anchored in the horizon, thus another possible
additional energy deposited into ejecta through the Blandford-Payne (BP) mechanism (Blandford \&
Payne 1982). The magnetic wind power can be estimated as follows (Livio et al. 1999, Meier et al.
2001),
\begin{equation}
L_{\rm bp} = (B_{\rm ms}^{\rm p})^2 r_{\rm ms}^4 \Omega_{\rm ms}^2 / 32c
\end{equation}
where $\Omega_{\rm ms}$ and $r_{\rm ms}$ are the Keplerian angular velocity and the radius of the
marginally stable orbit, respectively. The $r_{\rm ms}$ can be written as
\begin{equation}
r_{\rm ms}/ r_{\rm g} = 3 + Z_2 - [(3 - Z_1)(3 + Z_1 + 2Z_2)]^{1/2}
\end{equation}
where $r_{\rm g} \equiv {GM_{\bullet}/c^2}$, and $M_{\bullet}$ is the mass of BH. Here, $Z_1$ and
$Z_2$ are dependent on spin parameter of BH ($a_{\bullet}$). Namely, $Z_1 \equiv 1 + (1 -
a_{\bullet}^2)^{1/3} [(1 + a_{\bullet})^{1/3} + (1 - a_{\bullet})^{1/3}]$ for $0 \leq a_{\bullet}
\leq 1$, and $Z_2 \equiv (3 a_{\bullet}^2 + Z_1^2)^{1/2}$. So that, $\Omega_{\rm ms}$ is given by
\begin{equation}
\Omega_{\rm ms} = \left( \frac{GM_{\bullet}}{c^3} \right)^{-1} \frac{1}{\chi_{\rm ms}^3 + a_{\bullet}}
\end{equation}
where $\chi_{\rm ms} \equiv \sqrt{r_{\rm ms/ r_{\rm g}}}$. On the other hand, based on the
derivation of Blandford \& Payne (1982), the disk poloidal magnetic field $B_{\rm ms}^{\rm p}$ at
$r_{\rm ms}$ can be written as
\begin{equation}
B_{\rm ms}^{\rm p} = B_{\bullet}(r_{\rm ms} / r_{\bullet})^{-5/4}
\end{equation}
where $r_{\bullet} = r_{\rm g}(1 + \sqrt{1 - a_{\bullet}^2})$ and $B_{\bullet}$ are the horizon
radius of BH and the magnetic field strength threading the BH horizon, respectively (Lei et al.
2013; Liu et al. 2017 for a review). For more details, refer to Ma et al. (2018).

In our calculations, we assume that the distribution of ejecta mass ($M_{\rm ej}$) is from $10^{-4}
M_\odot$ to $10^{-2} M_\odot$, which can be accelerated to a relativistic speed ($\beta \sim 0.1 -
0.3$) by the magnetar wind, and the opacity is in the range of $\kappa \sim (0.1
- 10)\rm ~g\cdot cm^{-2}$. Also, $M_{\bullet}=3 M_{\odot}$ is adopted as the initial mass of
newborn BH, then, the total energy of the ejecta and shocked medium can be expressed as $E =
(\Gamma -1)M_{\rm ej}c^2 + \Gamma E_{\rm int}^{'} $,  where $\Gamma$ and $E_{\rm int}^{'}$ are the
Lorentz factor and the internal energy measured in the comoving frame, respectively. The energy
conservation is given as,
\begin{eqnarray}
d E_{\rm ej} = (L_{\rm inj} - L_{\rm e})dt
\end{eqnarray}
where $L_{\rm e}$ is the radiated bolometric luminosity. $L_{\rm inj}$ is the injection luminosity,
which is contributed by two terms. One is spin-down luminosity ($L_{\rm sd}$) from the magnetar and
radioactive power ($L_{\rm ra}$) before the magnetar collapsed into BH, and the second is magnetic
wind power ($L_{\rm bp}$) from a newborn BH after magnetar collapse. The mathematics formula can be
expressed as follows,
\begin{eqnarray}
L_{\rm inj} = \left\{ \begin{array}{ll}
\xi_1 L_{\rm sd} + L_{\rm ra}, & t \leq t_{col}\\
\xi_2 L_{\rm bp}, & t > t_{col}
\end{array}  \right.
\end{eqnarray}
Here, $\xi_1$ and $\xi_2 $ are corresponding to converted efficiencies from spin-down luminosity or
BP luminosity to the ejecta, respectively. A normal value of 0.3 is adopted with both $\xi_1$ and
$\xi_2 $ (Zhang \& Yan 2011). The $t_{\rm col}$ is characteristic collapse time from magnetar to
BH.

Together with the above considerations, the full dynamic evolution of the ejecta can be determined
by
\begin{equation}
\frac{d \Gamma}{dt} = \frac{L_{\rm inj} - L_{\rm e} - \Gamma \mathcal{D}(dE_{\rm int}^{'}/ dt^{'})}{M_{\rm
ej}c^2 + E_{\rm int}^{'}}
\label{dGamma}
\end{equation}
Here, we do not consider the energy loss, which is due to shock emission. $\mathcal{D} =
\Gamma+\sqrt{\Gamma^2-1}$ is the Doppler factor, and one can switch the comoving time ($dt^{'} $)
into observer's time ($dt$) based on the Doppler effect, $dt^{'} = \mathcal{D}dt$. The evolution of
internal energy in the comoving frame can be written as (Kasen \& Bildsten 2010, Yu et al. 2013)
\begin{eqnarray}
\frac{dE_{\rm int}^{'}}{dt^{'}} = L_{\rm inj}^{'} - L_{\rm e}^{'} - \mathcal{P}^{'}\frac{dV^{'}}{dt^{'}}
\label{dEint}
\end{eqnarray}
where the luminosity of radiative, spin-down of magnetar, and magnetic wind of a newborn BH in the
comoving frame are $L_{\rm sd}^{'} = L_{\rm sd}/\mathcal{D}^2$, $L_{\rm e}^{'} =
L_e/\mathcal{D}^2$, and $L_{\rm ra}/\mathcal{D}^2$ (Yu et al. 2013). The pressure $\mathcal{P}^{'}
= E_{\rm int}^{'}/3V^{'}$ is dominated by radiation, and the evolution of the comoving volume can
be addressed by
\begin{equation}
\frac{dV^{'}}{dt} = 4\pi R^2\beta c
\label{dV}
\end{equation}
where $\beta=\sqrt{1-\Gamma^{-2}}$. The the radius of ejecta ($R$) evolute with time as
\begin{equation}
\frac{dR}{dt} = \frac{\beta c}{(1-\beta)}
\label{dR}
\end{equation}
The radiate bolometric luminosity ($L_{\rm e}^{'}$) is read as (Kasen \& Bild- sten 2010; Kotera et
al. 2013)
\begin{eqnarray}
L_{\rm e}^{'} = \left\{ \begin{array}{ll}
\frac{E_{\rm int}^{'}c}{\tau R / \Gamma}, & t \leq t_{\tau}\\
\frac{E_{\rm int}^{'}c}{R/\Gamma}, &  t > t_{\tau}
\end{array}  \right.
\end{eqnarray}
where $\tau = \kappa(M_{ej}/V^{'})(R/\Gamma)$ is the optical depth of the ejecta, and $\kappa$ is
the opacity. $t_{\tau}$ is the time at which $\tau = 1$.

By assuming that the emission spectrum $\nu L_{\nu}$ is always characterized by the blackbody
radiation, and an effective temperature can be defined as (Yu et al. 2018)
\begin{equation}
T_{e}=(\frac{L_{e}}{4\pi R^{2}_{ph}\sigma_{SB}})^{1/4}
\label{temperature}
\end{equation}
where $\sigma_{\rm SB}$ is the Stephan-Boltzman constant, and $R_{\rm ph}$ is photospheric radius
when the optical depth equals unity in mass layer beyond. For given frequency $\nu$, the observed
flux density could be calculated as
\begin{equation}
F_{\nu}(t) = \frac{2\pi h \nu^3R^2}{D_{L}^{2}c^2}\frac{1}{\mathrm{exp}(h \nu
/kT_{e})-1}
\label{Fv}
\end{equation}
where $h$ is the Planck constant, and $D_{L}$ is the luminosity distance of the source. Finally, we
determine the monochromatic apparent magnitude by $M_\nu =
-2.5~\mathrm{log}_{10}\frac{F_\nu}{3631}\ \mathrm{Jy}$ (Yu et al. 2018).

\section{Sample Selection}
Since the stable magnetar signature typically invokes a plateau phase followed by a $t^{-2}$ decay
in X-ray emission (Zhang \& M{\'e}sz{\'a}ros 2001), but supra-massive NS signature require a
feature of X-ray emission with "internal plateau" (Troja et al. 2007; Rowlinson et al. 2010; L\"{u}
et al. 2015). We search for such a signature to decide how likely a GRB is to be powered by a
supra-massive NS, so that two criteria for our sample selection are adopted:
\begin{itemize}
\item We focus on short GRBs with internal plateau of X-ray emission\footnote{L\"{u} et al.
    (2015) performed a systematically searching for short GRBs sample with internal plateau by
    including the extended emission, which is extrapolated into X-ray band by assuming a single
    power-law spectrum, also see Rowlinson et al (2013). In this work, we only selected the
    short GRBs with internal plateau in X-ray emission, and do not consider the internal
    plateau which caused by extrapolating from BAT to X-ray.}, which may be the signature of
    supra-massive NS as the central engine and suffering a collapse process from supra-massive
    NS to new-born BH.
\item In order to obtain quantitatively fitting of the data and more intrinsic information of
    GRBs, we focus on bursts with redshift measurement.
\end{itemize}

Following the above criteria, we systematically search for 122 short GRBs observed by the {\em
Swift} Burst Alert Telescope (BAT) from the launch of {\em Swift} to 2020 September, and 105 short
GRBs have the X-ray observations with more than five data points. The X-ray data of those GRBs are
taken from the {\em Swift} X-ray Telescope (XRT) website\footnote{$\rm
https://www.swift.ac.uk/xrt\_curves/allcurves.php$} (Evans et al. 2009). We then perform a temporal
fit to the X-ray light curve with a smooth broken power law in the rest frame,
\begin{eqnarray}
F = F_0 \left[ \left(\frac{t}{t_b}\right)^{\omega\alpha_1} + \left( \frac{t}{t_b}\right)^{\omega\alpha_2}
\right ]^{-1/\omega}
\end{eqnarray}
Here, $t_{\rm b}$ is the break time, $F_{\rm b} = F_0 \cdot 2^{-1/\omega}$ is the flux at $t_{\rm
b}$, and $\alpha_1$ and $\alpha_2$ are decay indices before and after the break, respectively.
$\omega$ describe the sharpness of the break, and the larger $\omega$ describes the sharper break.
In order to identify a possible an internal plateau, $\omega=10$ is adopted in our analysis (Liang
et al. 2007). We find that 17 bursts exhibit an internal plateau. Within this sample, only five
bursts (e.g., GRBs 060801, 090515, 100625A, 101219A, and 160821B) have the redshift measured. The
fitting results of those bursts are presented in Figure 1 and Table 1. Since the GRBs 060801,
090515, 100625A, and 101219A are reported  in Rowlinson et al. (2013) and L\"{u} et al. (2015), and
the internal plateau feature of GRB 160821B is also identified in L\"{u} et al. (2017) and Troja et
al. (2019). Moreover, even the decay slope following the plateau of X-ray emission for GRB 090510
is steeper than 2, and measured redshift, but the spectrum of X-ray and optical during the plateau
phase is consistent with a standard forward shock model (Ackermann et al. 2010; De Pasquale et
al.2010; Pelassa \& Ohno 2010; Nicuesa Guelbenzu et al. 2012). So this case is not included in our
sample.

\section{Detectability of ``Merger-nova" emission}
It is interesting to compare the brightness of $r$-process powered kilonovae, magnetar-powered
merger-novae, and the BH magnetic wind powered merger-novae claimed in the literature. How do the
brightness of the merger-nova emission depend on those three energy sources, e.g.,
$r$-process-powered (marked $S1$), spin energy from supramassive NS (marked $S2$), and magnetic
wind from the accretion disk of a newborn BH (marked $S3$). Figure \ref{fig:mergernova} shows the
numerical calculation of merger-nova emission in the $R$ band by considering only $S1$, $S1+S2$,
and $S1+S2+S3$ to adopt the typical value of model parameters. We find that a significant
contribution from $S2$ and $S3$ for merger-nova emission occurs by comparing the contribution of
only $S1$.

Based on the fits of the internal plateau of X-ray emission, one can roughly estimate the initial
spin-down luminosity $L_0\sim 4\pi D_{L}^2 F_0$ which is also presented in Table 1 for our sample.
In order to infer the power of spin-down of a magnetar and the magnetic wind of a newborn BH, we
roughly set up $\alpha=\alpha_2$, and $\tau=t_{\rm col}=t_{b}$. By combining with those three
energy sources of merger-nova emission, Figure \ref{GRB060801}-\ref{GRB160821B} show the possible
merger-nova light curve of our sample in $K$-, $r$-, and $U$-bands with variations of $M_{\rm ej}$,
$\kappa$, and $\beta$. Moreover, we also overplot the upper limit detected of instruments for we
also overplot the upper limit detected of instruments for Vera C. Rubin, ZTF, Pan-STARRS, and
Roman. In order to compare the real optical observations of those candidates with the model, we
collect the optical or near-infrared (NIR) data that was observed by optical telescopes of them as
much as possible if the optical data are available or even upper limits.

The top three panels of Figure \ref{GRB060801} show the merger-nova light curve with variation of
$M_{\rm ej}=10^{-2} M_{\odot}$, $10^{-3} M_{\odot}$, and $10^{-4} M_{\odot}$ for fixed
$\kappa=0.1~\rm cm^{2}~g^{-1}$ and $\beta=0.1$. In the middle three panels of Figure
\ref{GRB060801} also show its light curve, but fixed $M_{\rm ej}=10^{-2} M_{\odot}$ and
$\kappa=0.1~\rm cm^{2}~g^{-1}$ with variation of $\beta=$0.1, 0.2, 0.3. Bottom three panels of
Figure \ref{GRB060801} present the light curve for fixed $M_{\rm ej}=10^{-2} M_{\odot}$ and
$\beta=$0.1 with a variation of $\kappa=0.1$, $1.0$, and $10~\rm cm^{2}~g^{-1}$. It is clear to see
that a lower ejecta mass and higher ejecta velocity would power a brighter merger-nova emission,
but the variation of opacity is not significant enough to effect the brightness of merger-nova.
Moreover, the numerical calculation shows that the merger-nova emission of GRB 060801 in $K$, $r$,
and $U$ bands with variation of $M_{\rm ej}$, $\kappa$, and $\beta$ is very difficult to detect
using all of the above instruments, except for the case of large ejecta mass $M_{\rm ej}=10^{-2}
M_{\odot}$. Similar results are also shown for GRB 090515 in Figure \ref{GRB090515}.

Following the same method as in Figure \ref{GRB060801}, we also present the merger-nova emission
light curves of the other four GRBs (e.g., GRBs 090515, 100625A, 101219A, and 160821B) in our
sample from Figure \ref{GRB090515} to Figure \ref{GRB160821B}. We are very hopeful to detect the
merger-nova emission of GRBs 090515, 100625A, and 101219A using the more sensitive instruments of
Roman, Vera C.Rubin, and Pan-STARRS, but it seems to be difficult to detect them using ZTF. Figure
\ref{GRB160821B} shows the possible merger-nova emission of the nearby short GRB 160821B. One can
see that it is bright enough to be detected by all Vera C. Rubin, Pan-STARRS, ZTF, as well as Roman
for given model parameters. We will discuss more details of this case at the end of Section 5.

\section{Conclusion and discussion}
The observed X-ray emission internal plateau in some short GRBs suggests that one possible remnant
of double NS mergers is a supramassive NS that is supported by rigid rotation and survives for
hundreds of seconds before collapsing into a BH. An optical/infrared transient (e.g., merger-nova)
is generated from the ejected materials and powered by radioactive decay from the $r$-process, spin
energy from supramassive NS, and magnetic wind from the accretion disk of newborn BH. In this work,
we systematically search for the signature of a supramassive NS central engine by analyzing the
X-ray emission of short GRBs, and calculate how bright this merger-nova is for the given model
parameters. Five candidates of short GRBs (GRBs 060801, 090515, 100625A, 101219A, and 160821B) are
found to be carrying such a feature with redshift measurements. By detail numerical calculations,
we find that the merger-nova emission of GRB 060801 in $K$, $r$, and $U$ bands with variations of
$M_{\rm ej}$ ($10^{-4}-10^{-2} M_{\odot}$), $\kappa$ ($0.1-10 ~\rm cm^{2}~g^{-1}$), and $\beta$
($0.1-0.3$) is very difficult to be detected by Vera C. Rubin, Pan-STARRS, as well as ZTF, except
for the case of less ejecta mass $M_{\rm ej}=10^{-4} M_{\odot}$. However, it is very hopeful to
detect the merger-nova emission of GRBs 100625A, and 101219A by more sensitive instruments of
Roman, Vera C. Rubin and Pan-STARRS, but seems to be difficult detected by ZTF.

For GRB 090515, there are two real optical data in $r$ band observed by the optical telescope (Fong
\& Berger 2013), and the model prediction of the merger-nova signal for given model parameters are
higher than those two real optical data. However, no any evidence was found for recent radio
surveys of short GRBs (Klose et al. 2019; Ricci et al. 2021). This contradiction between
observations and model prediction may be caused by the unreasonable selected model parameters
(e.g., higher ejecta mass, lower opacity, and higher speed). More interestingly, the merger-nova
emission of the nearby short GRB 160821B is indeed detected or even some upper limits by optical
telescopes (Kasliwal et al. 2017, called it kilonova), and the presence of a merger-nova was also
discussed by Kasliwal et al. (2017), Jin et al. (2018), and Troja et al. (2019). The radio surveys
of this case are also reported by Ricci et al. (2021). In our calculations, the merger-nova
emission of this case is bright enough to be detected by Roman, Vera C. Rubin, Pan-STARRS, and ZTF,
and it is consistent with current observed data of merger-nova. Recently, Ma et al. (2020) proposed
that the merger-nova of this case is possibly powered through the Blandford-Znajek mechanism (BZ;
Blandford \& Znajek 1977) forming a newborn BH, and the observed data of merger-nova is consistent
with the physical process from the magnetar collapsing into a BH. In order to confirm the real
model parameters of this case, we are working on the Markov Chain Monte Carlo (MCMC) method to
present the values of model parameters in another paper (in preparation) of this case, as well as
other short GRBs if the optical data are good enough (An et al. 2021, in preparation).

The criteria for our sample selection invoked the X-ray internal-plateau emission, which indicates
the signature of supramassive NS collapse into BH. Alternatively, Yu et al. (2018) proposed that
the internal-plateau emission in some short GRBs may be caused by an abrupt suppression of the
magnetic dipole radiation of a remnant NS. If this is the case, the energy injection from NS will
be contributed to the merger-nova at a later time. In this paper, we do not consider this effect.
In any case, in order to investigate the physical process of the magnetar central engine in short
GRBs, we therefore encourage intense multiband optical follow-up observations of short GRBs with
internal plateaus to catch such merger-nova signatures in the future. More importantly, the
observed kilonova or merger-nova together with the GW signal from a double NS merger and an NS-BH
merger will provide a good probe to study the accurate Hubble constant, the equation of state of an
NS, as well as the origin of heavier elements in the universe. With the improvement of detection
technology, we are looking forward to observing more and more GW events associated with merger-nova
from compact stars in the future.

\acknowledgments We are very grateful to thank the referee for constructive comments and
suggestions. We acknowledge the use of the public data from the {\em Swift} data archive and the UK
{\em Swift} Science Data Center. We would like to thank Yunwei Yu for sharing his original code
used to produce merger-nova, and we also thank Dabin Lin, Jia Ren, Shenshi Du, and Cheng Deng for
helpful discussions. This work is supported by the National Natural Science Foundation of China
(grant Nos. 11922301, 11851304, and 11533003), the Guangxi Science Foundation (grant Nos.
2017GXNSFFA198008, 2018GXNSFGA281007, and AD17129006), the Program of Bagui Young Scholars Program
(LHJ), and special funding for Guangxi distinguished professors (Bagui Yingcai and Bagui Xuezhe).

\newpage
\begin{center}
\begin{deluxetable}{cccccccccccccc}
\tablewidth{0pt} \tabletypesize{\footnotesize}
\tablecaption{Observed properties of short GRBs in our sample} \tablenum{1}

\tablehead{ \colhead{GRB}& \colhead{$z$\tablenotemark{a}}& \colhead{$T_{90}$} &
\colhead{$L_{0,49}$\tablenotemark{b}} & \colhead{$\alpha_1$\tablenotemark{c}} &
\colhead{$\alpha_2$\tablenotemark{c}} & \colhead{$t_b$\tablenotemark{c}}\\
\hline \colhead{Name}& \colhead{}& \colhead{(s)} & \colhead{($\rm erg~s^{-1}$)} & \colhead{}&
\colhead{} & \colhead{(s)}} \startdata
060801    &  1.13   & 0.5     & 0.73 $\pm$ 0.07   & 0.1 $\pm$ 0.11   & 4.35 $\pm$0.26   & 212 $\pm$11 \\
090515    &  0.4    & 0.036   & 1.24 $\pm$ 0.05   & 0.10$\pm$ 0.08   &12.62 $\pm$0.5    & 178$\pm$3 \\
100625A   &  0.425  & 0.33    & 0.042$\pm$0.03    & 0.26$\pm$ 0.44   & 2.47 $\pm$0.18   & 200 $\pm$41 \\
101219A   &  0.718  & 0.6     & 0.56 $\pm$0.05    & 0.13$\pm$ 0.19   & 20.52$\pm$8.01   & 197 $\pm$10 \\
160821B   &  0.16   & 0.48    & 0.18 $\pm$0.06    & 0.21$\pm$ 0.14   & 4.52 $\pm$0.45   & 180 $\pm$46 \\
\enddata
\tablenotetext{a}{The measured redshifts are taken from the published papers and GCNs.}
\tablenotetext{b}{The plateau luminosity of X-ray emission based on the fits.}
\tablenotetext{c}{$\alpha_1$ and $\alpha_2$ are the decay slopes before and after the break time,
and $t_{\rm b}$ is the break time of the light curves from our fitting.} \tablerefs{Redshift
reference for our sample. GRB 060801-$z$: Cucchiara et al. (2006); GRB 090515-$z$: Fong \& Berger
(2013); GRB 100625A-$z$: Fong et al. (2013); GRB101219A-$z$: Fong et al. (2013); GRB 160821B-$z$:
Troja et al. (2019).}
\end{deluxetable}
\end{center}


\clearpage
\begin{figure}
\centering
\includegraphics [angle=0,scale=0.25] {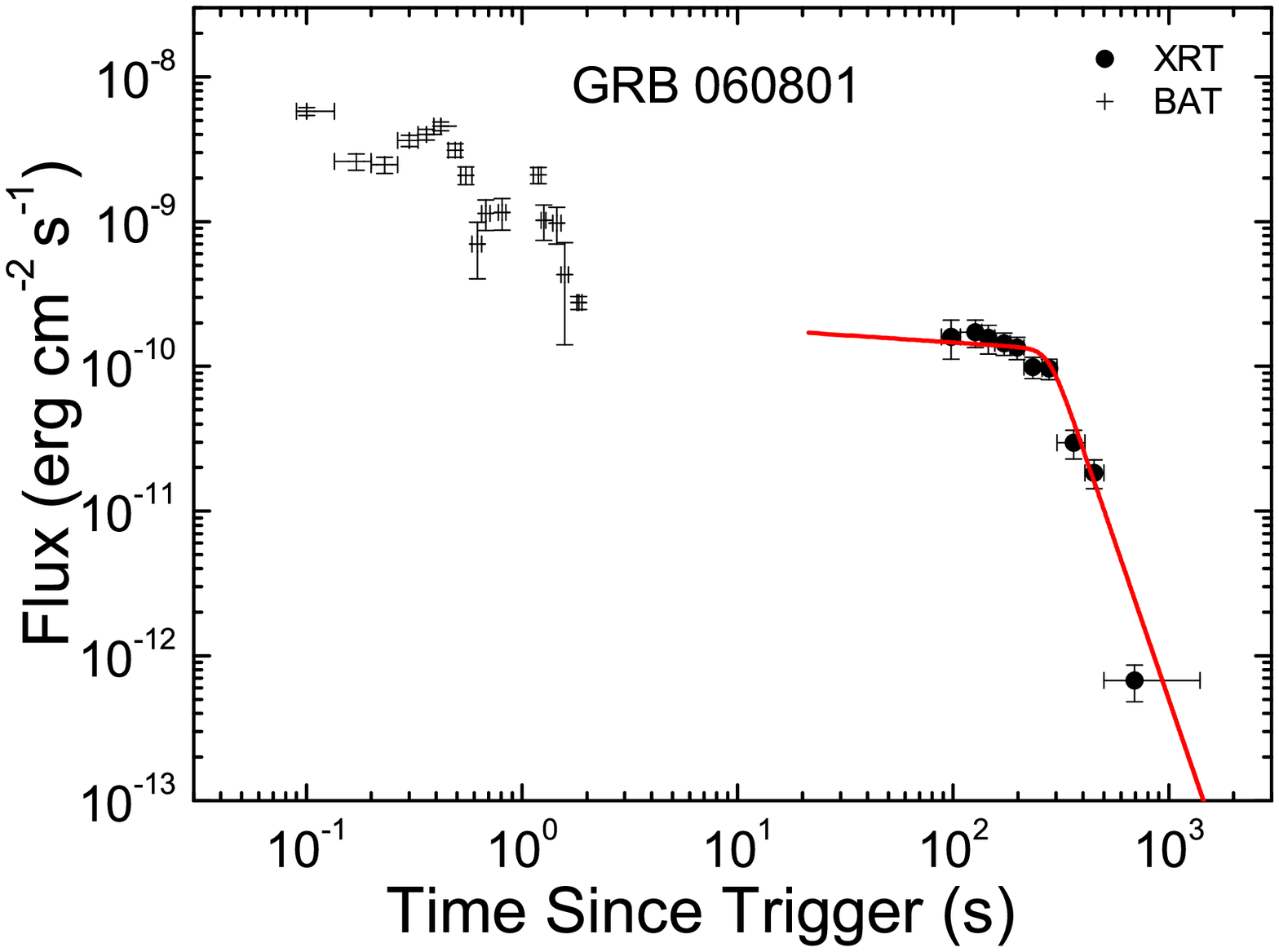}
\includegraphics [angle=0,scale=0.25] {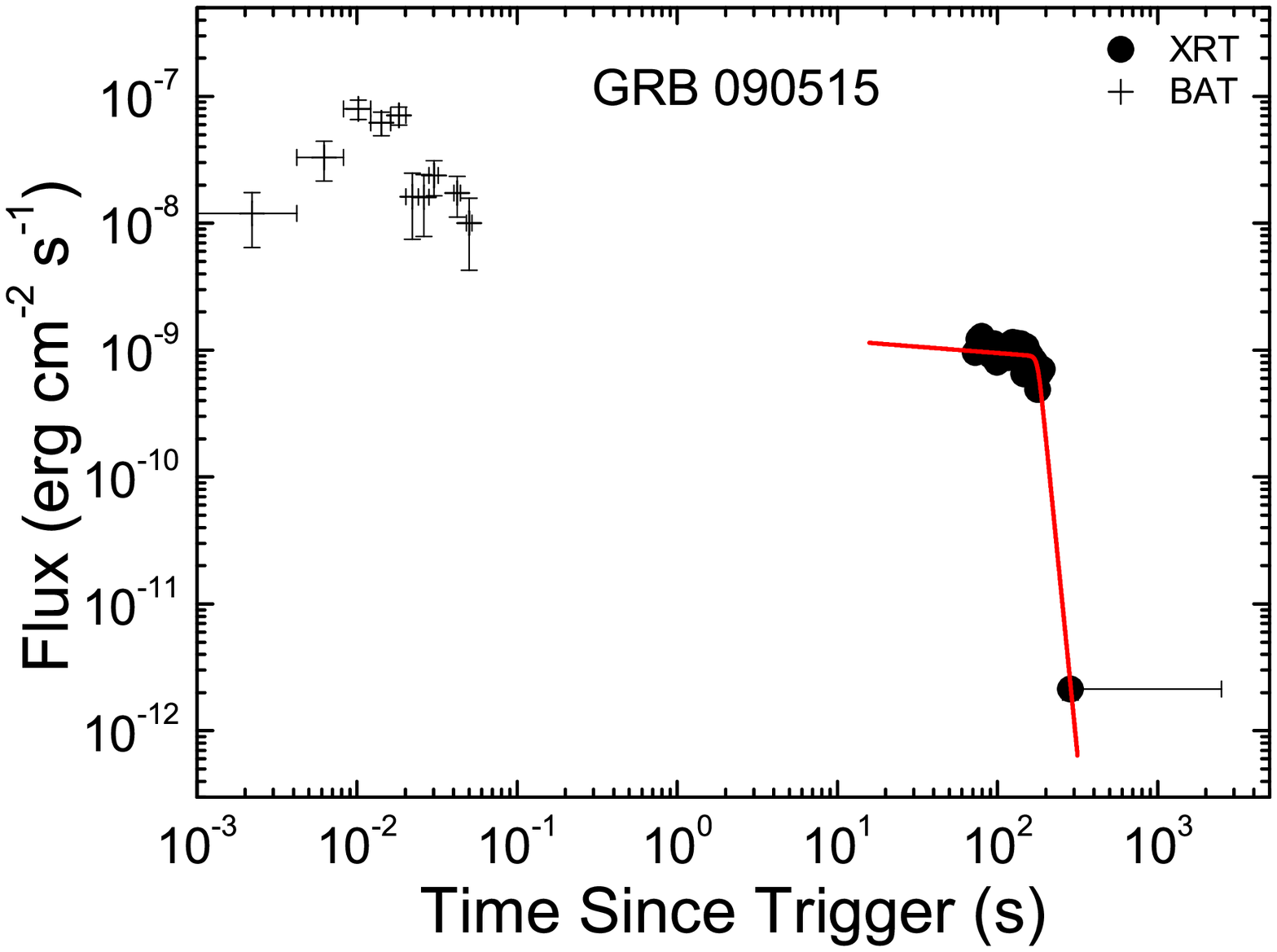}
\includegraphics [angle=0,scale=0.25] {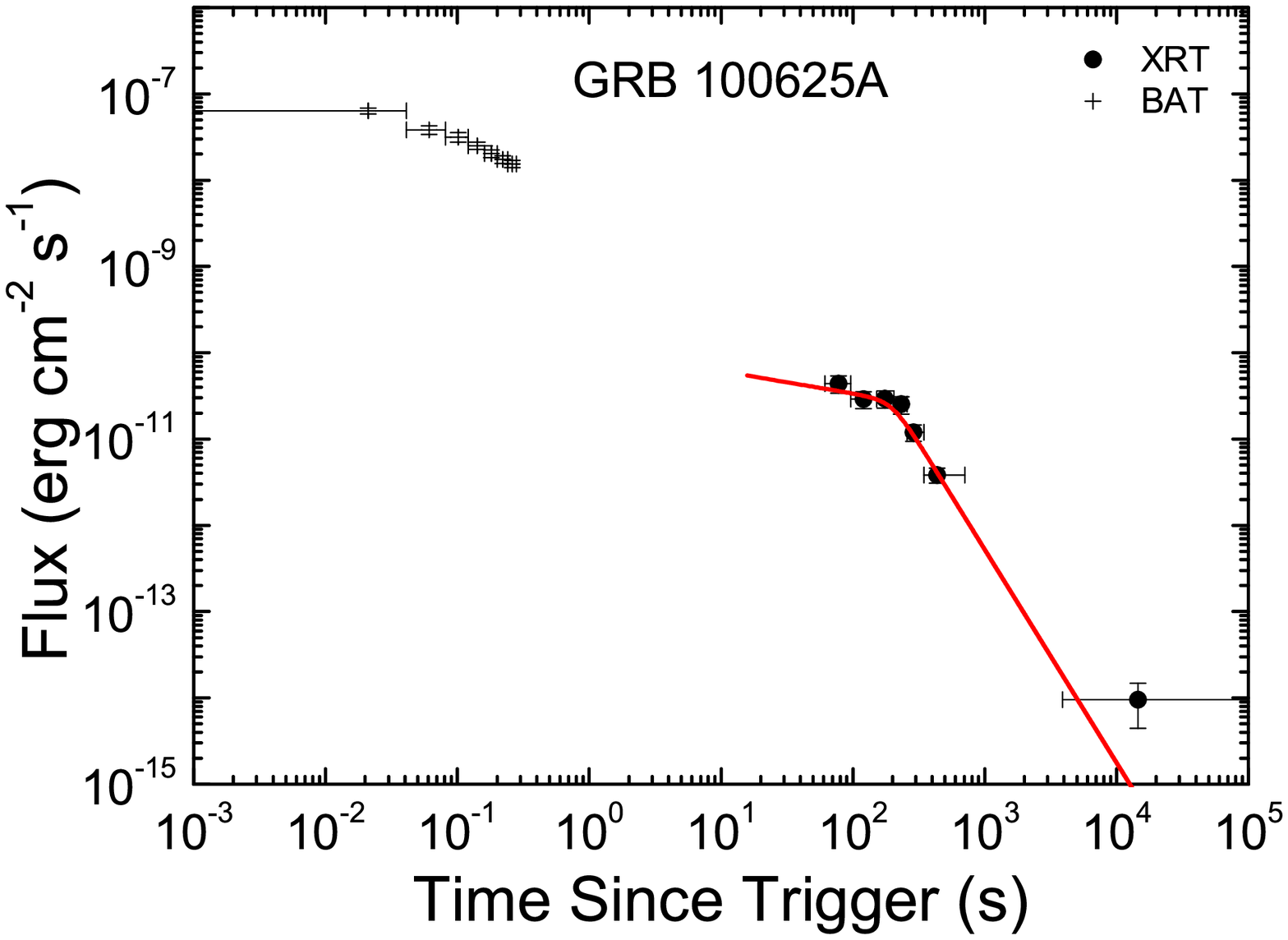}
\includegraphics [angle=0,scale=0.25] {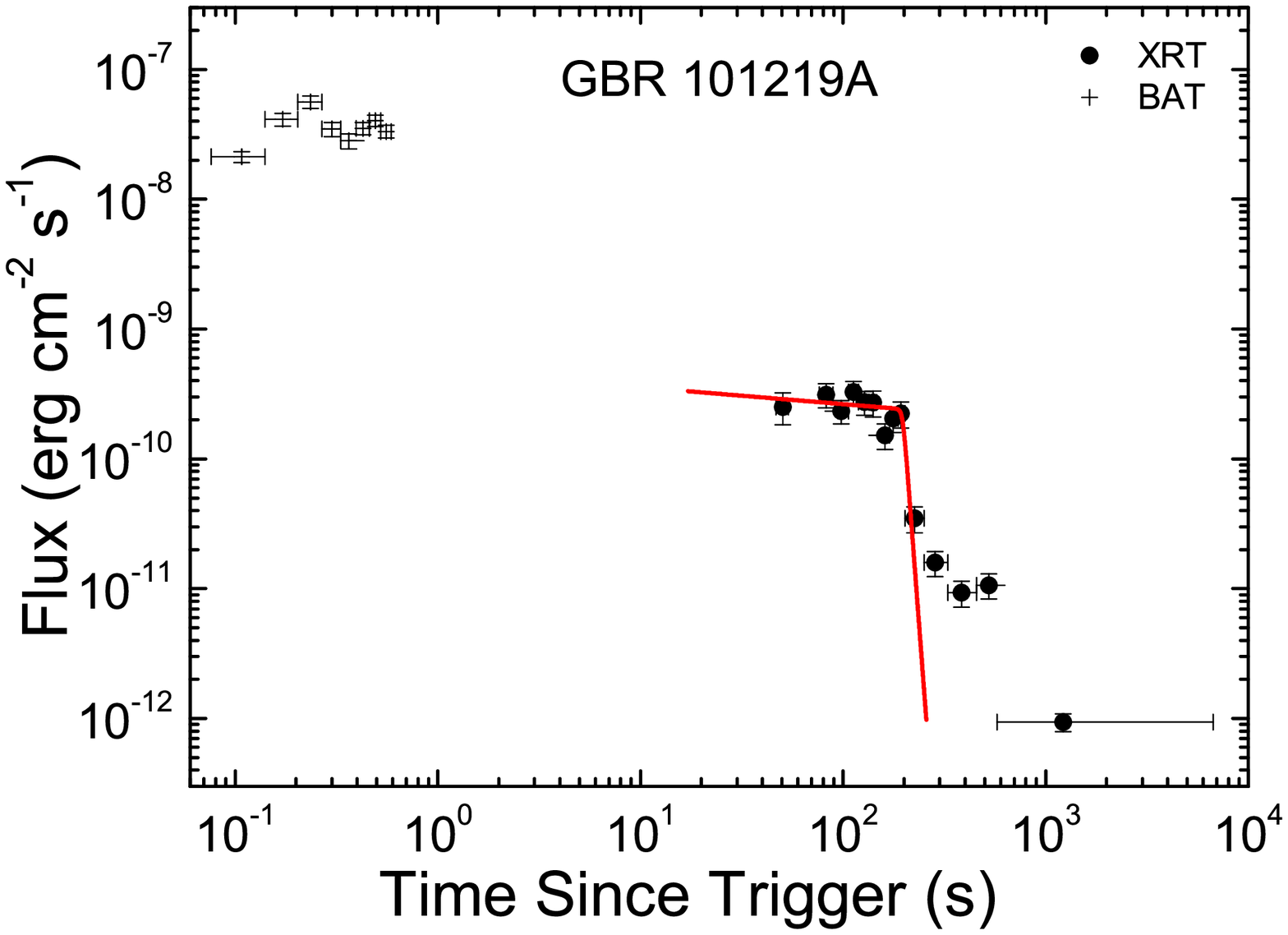}
\includegraphics [angle=0,scale=0.25] {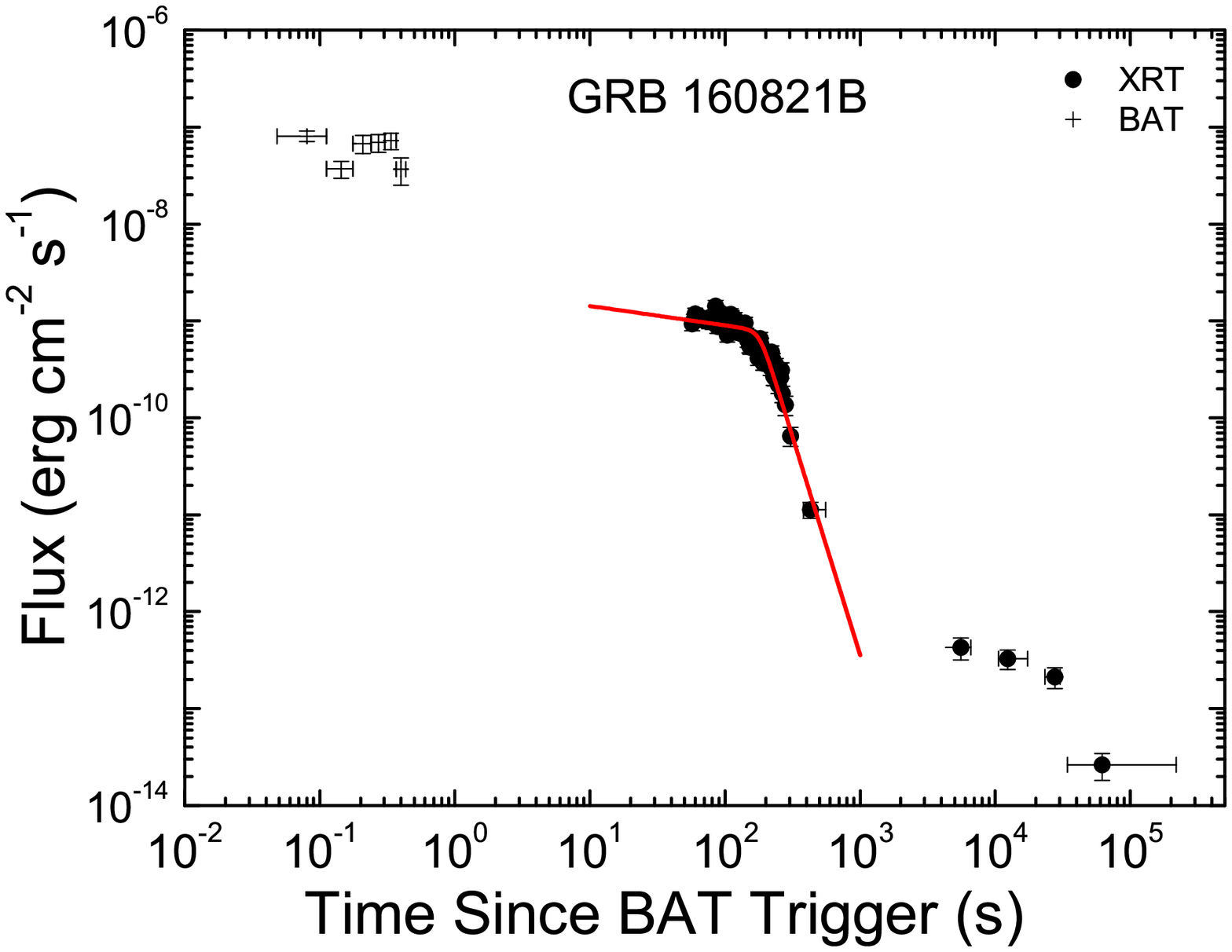}
\caption{BAT-XRT rest-frame light curves of the GRBs in our internal sample.
Black pluses are BAT data extrapolated to the XRT band, and black points (with
error bars) are the XRT data. The red solid curves are the best fits with a
smooth broken power-law model to the data.}
\label{fig:LC}
\end{figure}

\begin{figure}
\centering
\includegraphics [angle=0,scale=0.6] {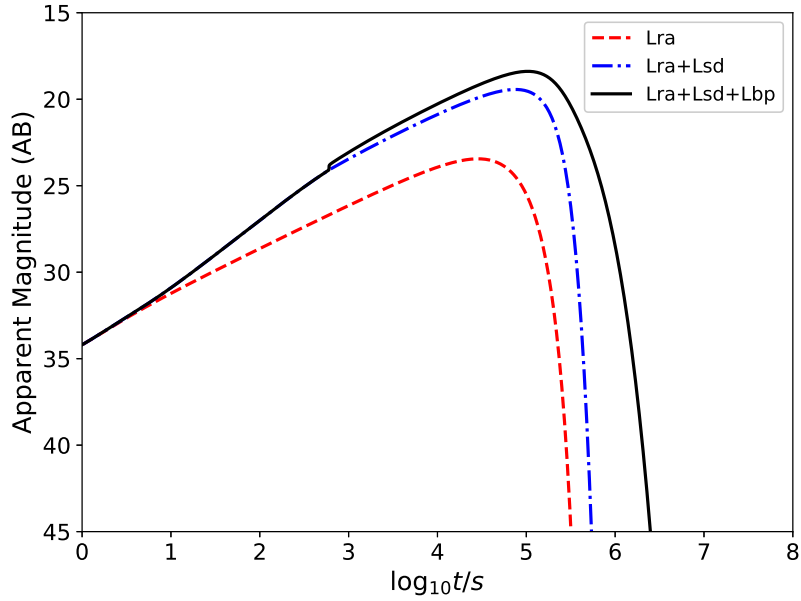}
\caption{Merger-nova light curve of the $r$ band for considering the distributions of three energy sources.
(1) Red dashed line is only considering the radioactive decay of heavier elements (marked as $S1$) at $D_{L}=400$
Mpc.
(2) Blue dashed-dotted line is the contributions from both $S1$ and spin-down
energy of a supramassive NS (marked as $S2$) with typical values $M_{\rm ej}=0.1 M_{\odot}$, $\beta=0.1$,
$\kappa=1.0~\rm cm^{2}~g^{-1}$, $\alpha=5$, $\tau=t_{\rm col}=600$ s, and
$L_0=1.0 \times 10^{49}~{\rm erg~s^{-1}}$. (3) Black solid line is the sum of $S1$, $S2$, and magnetic wind
power of a newborn BH (marked as $S3$) with $M_{\bullet}=3 M_{\odot}$, $M_{d}=0.1 M_{\odot}$
(initial mass of BH), and viscosity coefficient $\hat{\alpha}=0.1$.}
\label{fig:mergernova}
\end{figure}
\begin{figure}
\centering
\includegraphics[width=.25\textwidth]{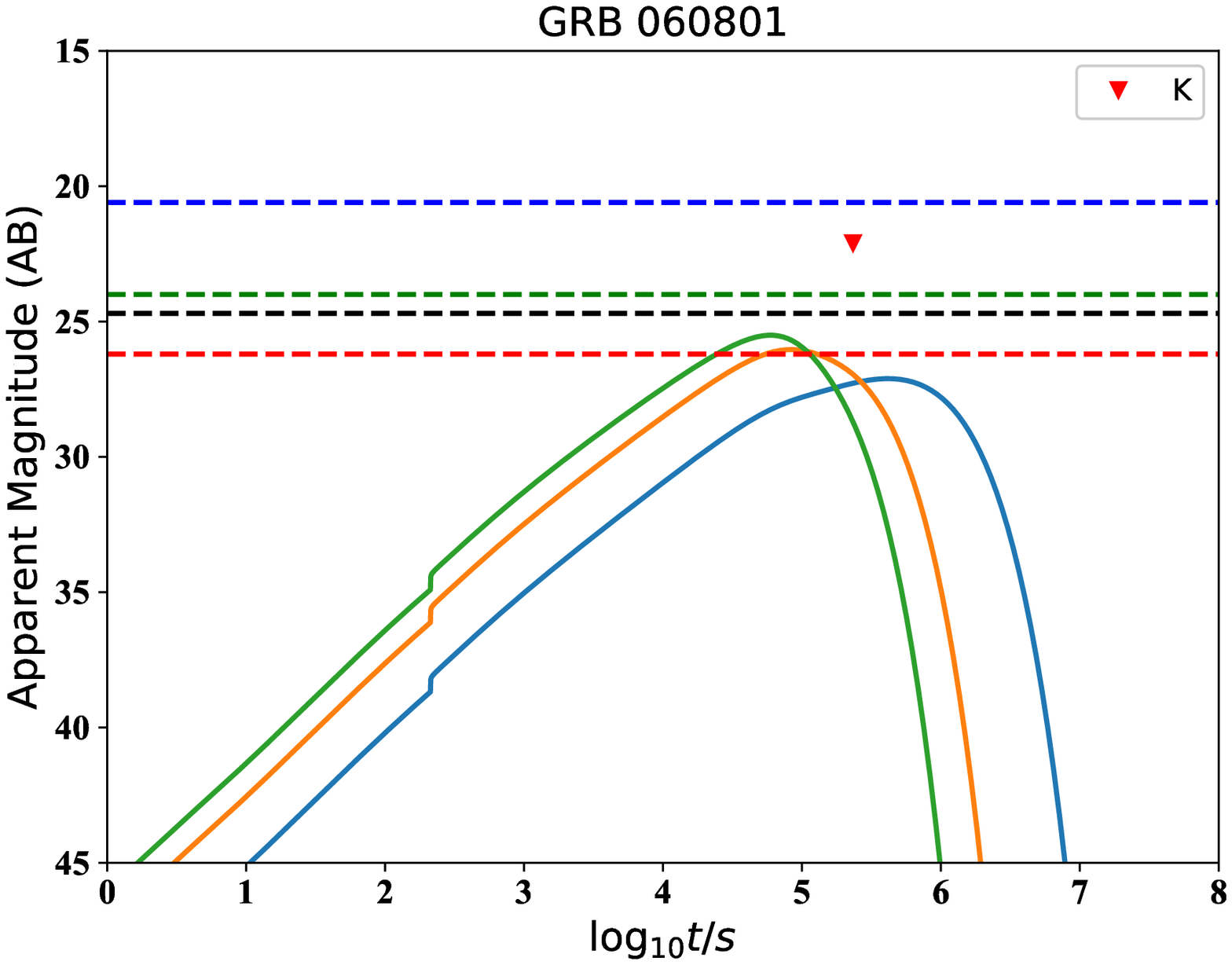}
\includegraphics[width=.25\textwidth]{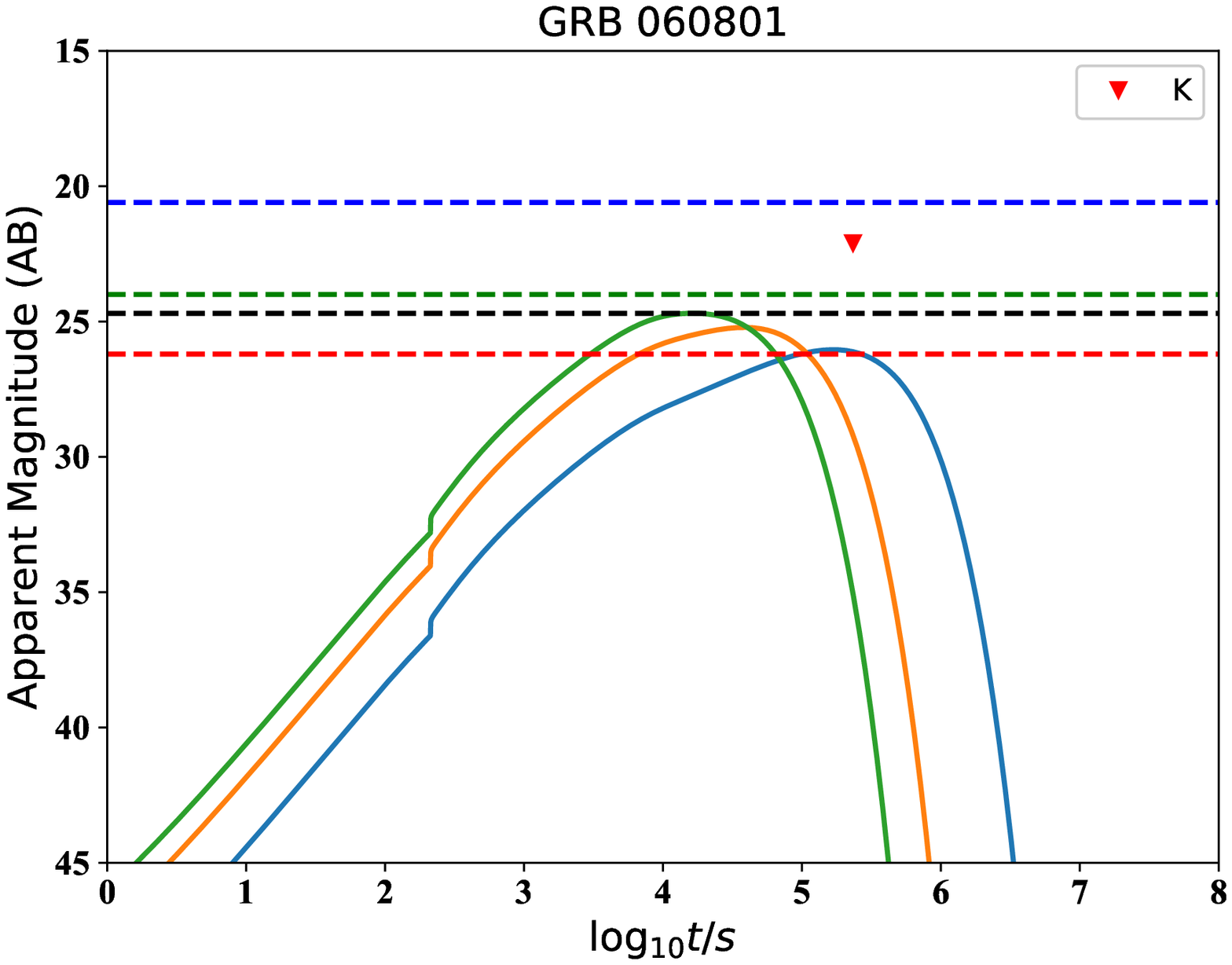}
\includegraphics[width=.25\textwidth]{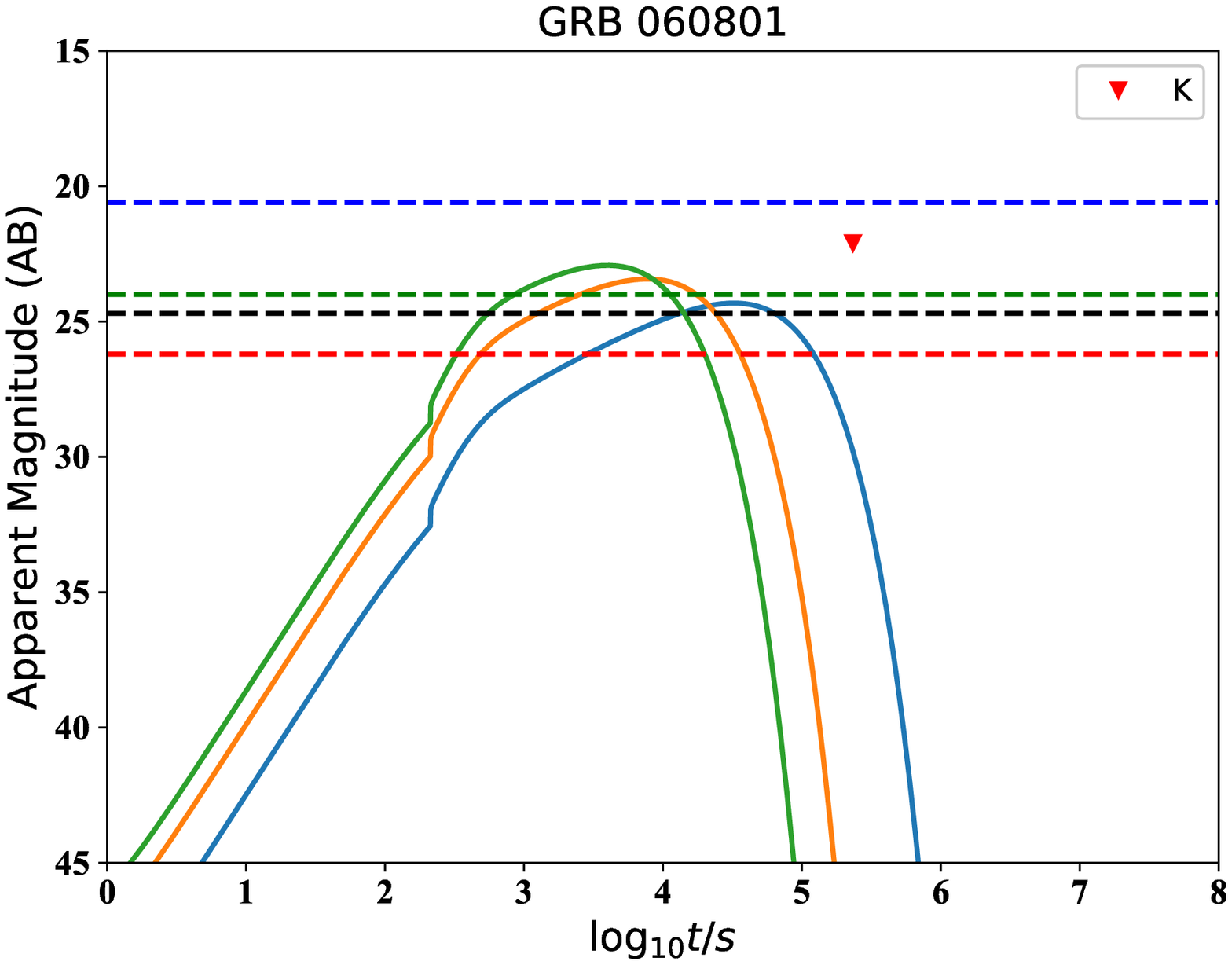}
\includegraphics[width=.25\textwidth]{GRB060801M001B01A01.eps}
\includegraphics[width=.25\textwidth]{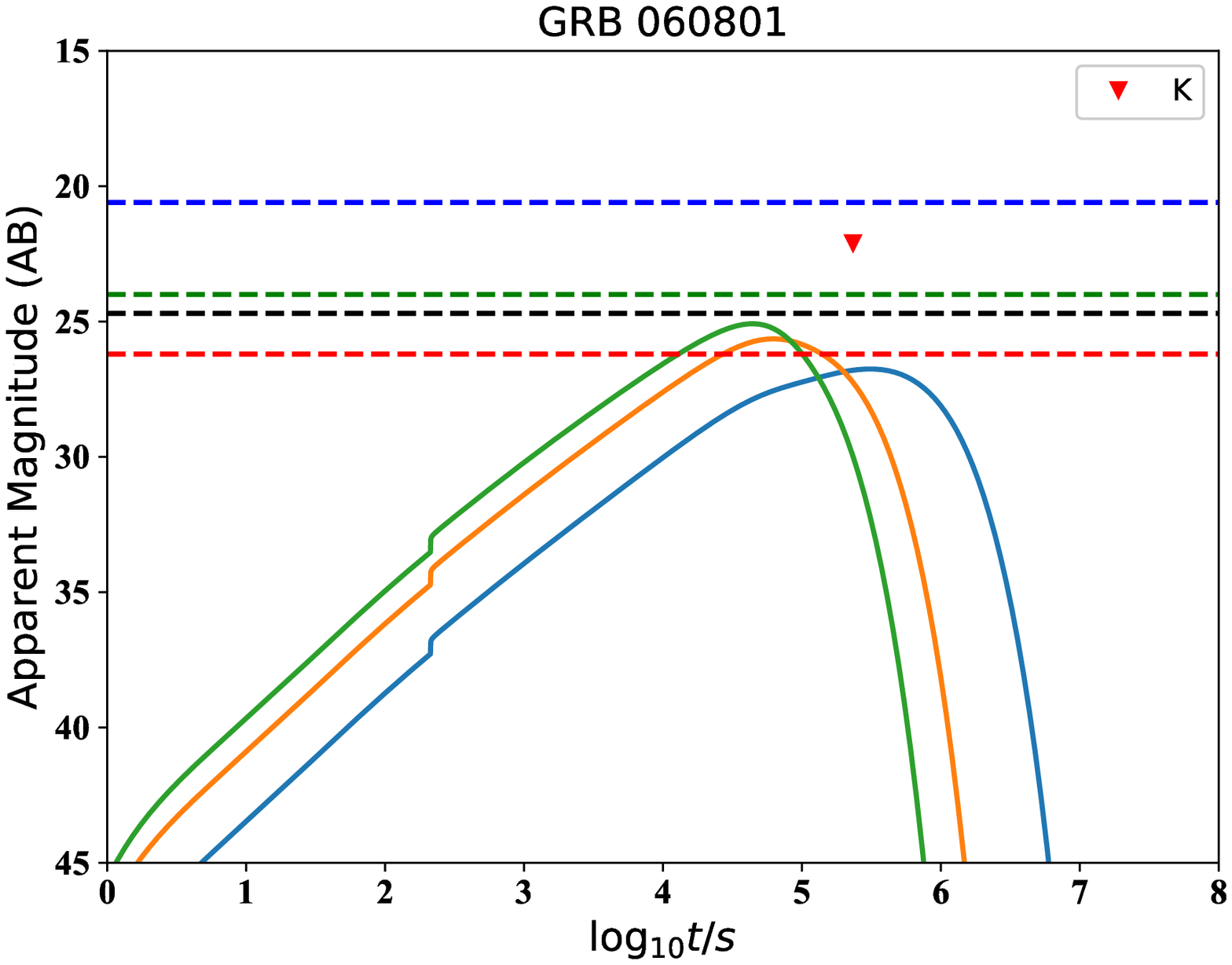}
\includegraphics[width=.25\textwidth]{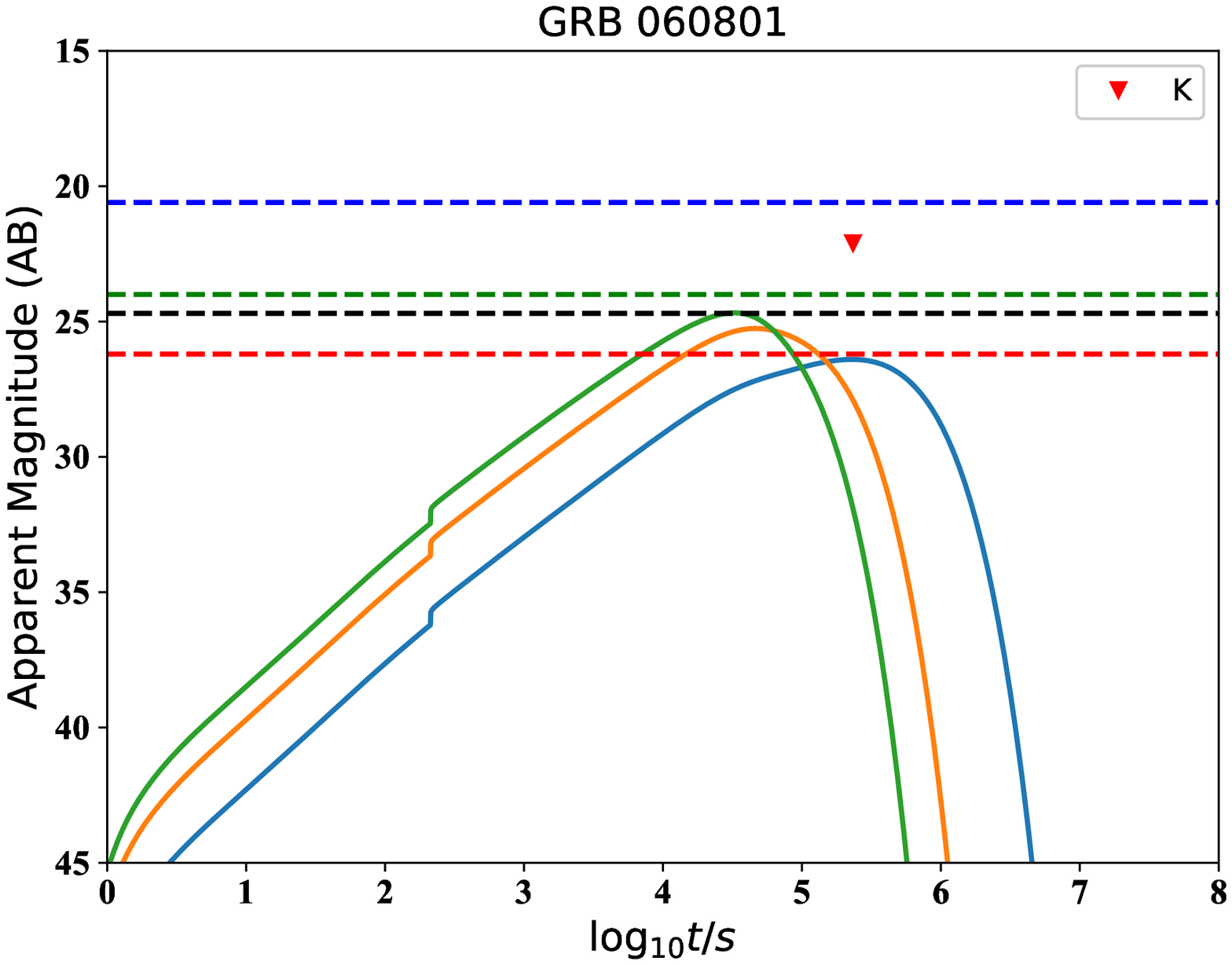}
\includegraphics[width=.25\textwidth]{GRB060801M001B01A01.eps}
\includegraphics[width=.25\textwidth]{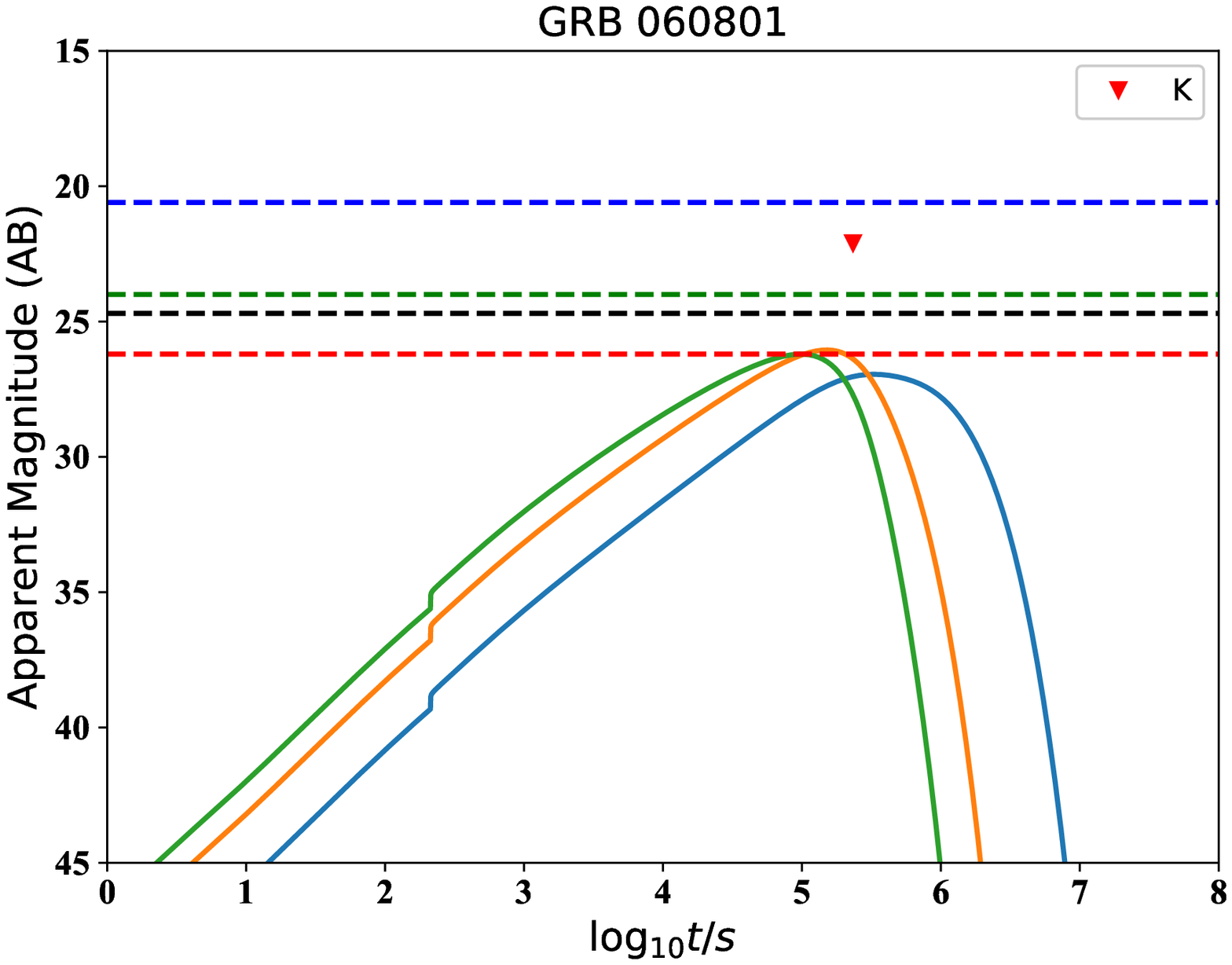}
\includegraphics[width=.25\textwidth]{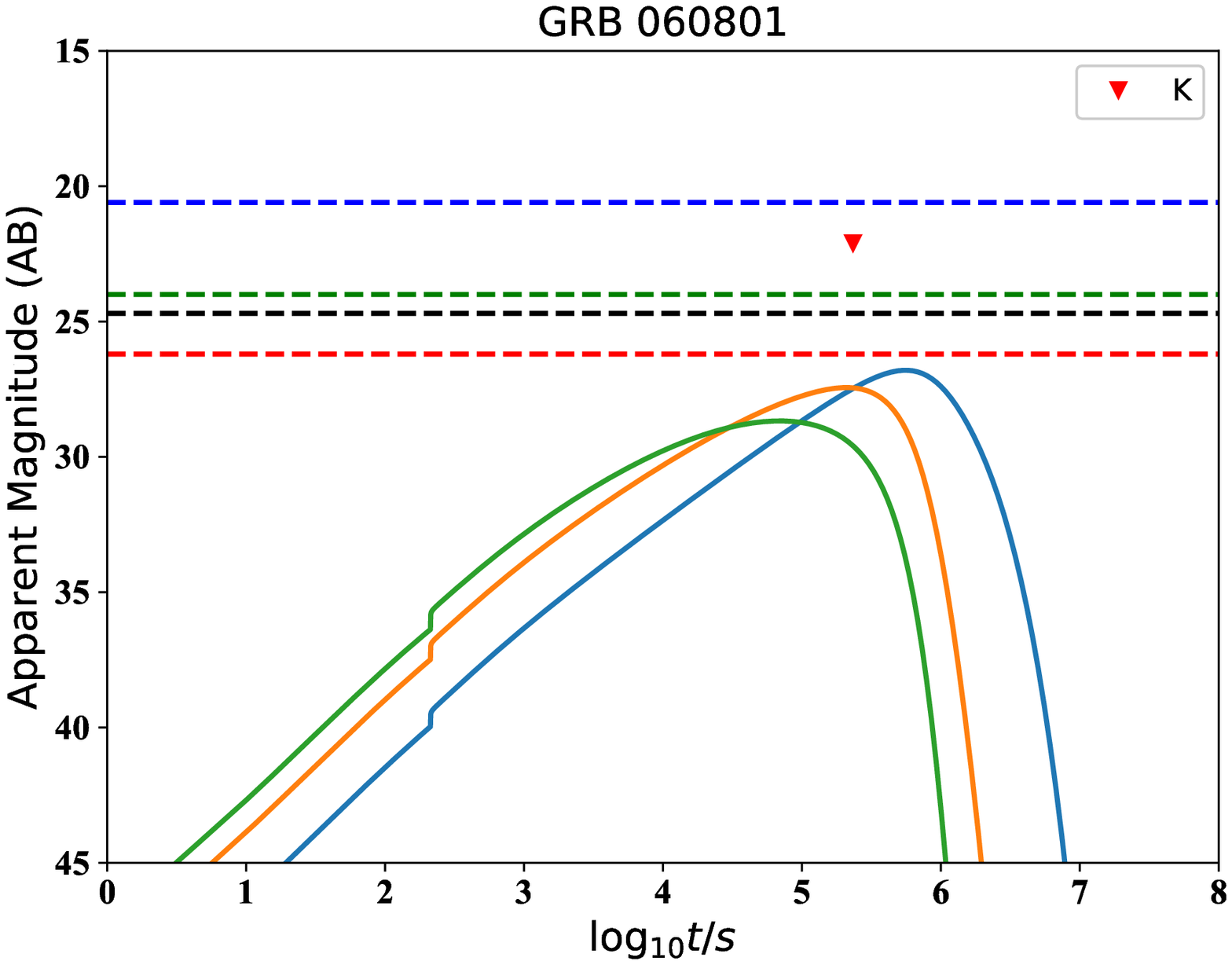}
\caption{Merger-nova light curve in $K$ (blue solid line), $r$ (orange solid line), and $U$ bands
(green solid line) of GRB 060801 by considering $S1$, $S2$, and $S3$. ({\em Top panels}):
we fixed $\kappa=0.1~\rm cm^{2}~g^{-1}$ and $\beta=0.1$, adopting a variation of
$M_{\rm ej}=10^{-4} M_\odot, 10^{-3} M_\odot$, and $10^{-2} M_\odot$, respectively.
({\em Middle panels}): similar to top panels, but with fixed $M_{\rm ej} = 0.01 M_{\odot}$ and
$\kappa=0.1~\rm cm^{2}~g^{-1}$, taking a variation of $\beta=$0.1, 0.2, 0.3.
({\em Bottom panels}): fixed $M_{\rm ej} = 0.01M_{\odot}$ and $\beta=0.1$, and variation of
$\kappa=$0.1, 1.0, and 10$~\rm cm^{2}~g^{-1}$.
The black dotted, blue dotted, green dotted, and red dotted lines correspond to the observed limitation
of Vera C. Rubin, ZTF, Pan-STARRS, and Roman, respectively. The observed upper limit of optical data
(red triangles) in the $K$ band are taken from Piranomonte et al. (2006).}
\label{GRB060801}
\end{figure}

\begin{figure}
\centering
\includegraphics[width=.25\textwidth]{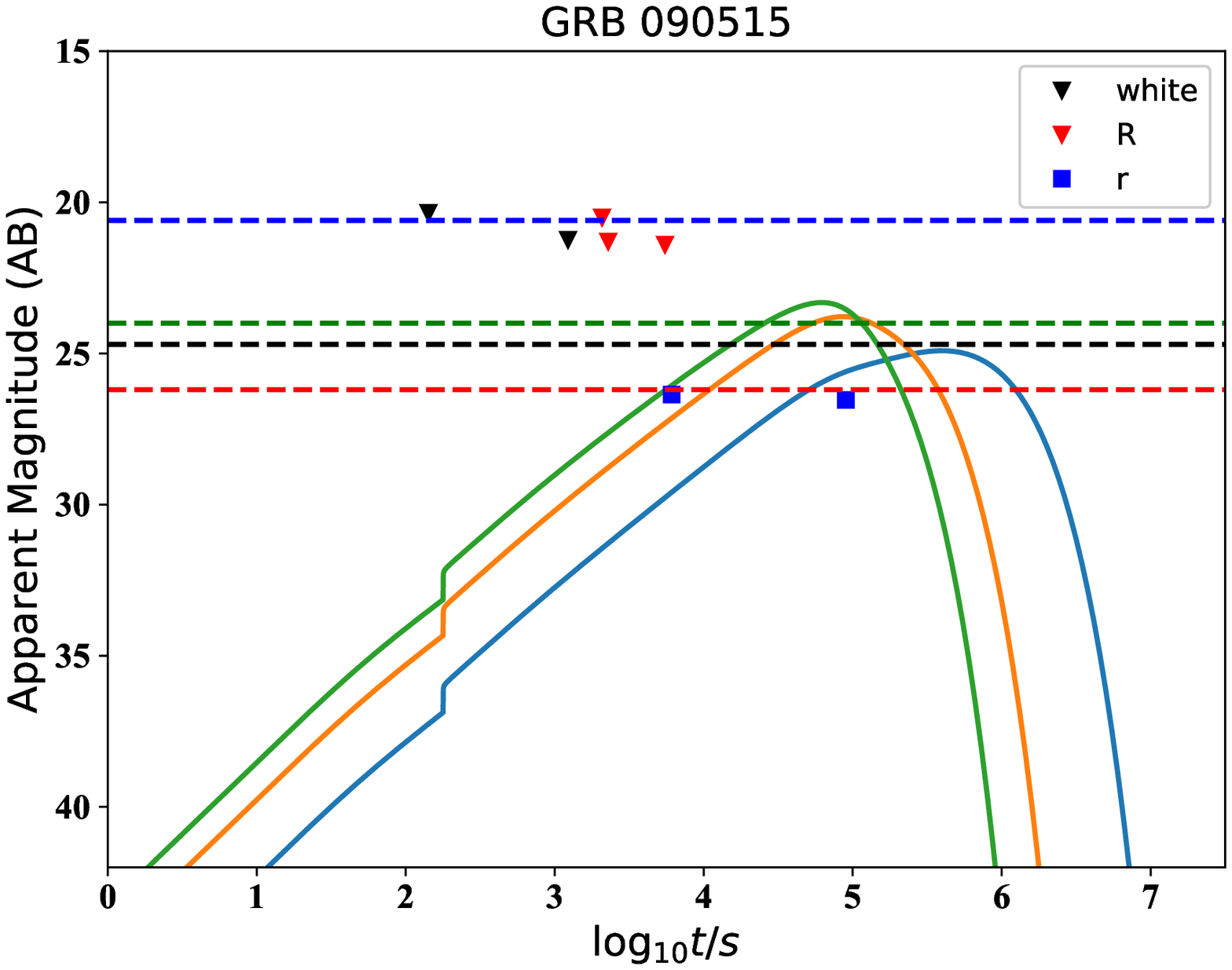}
\includegraphics[width=.25\textwidth]{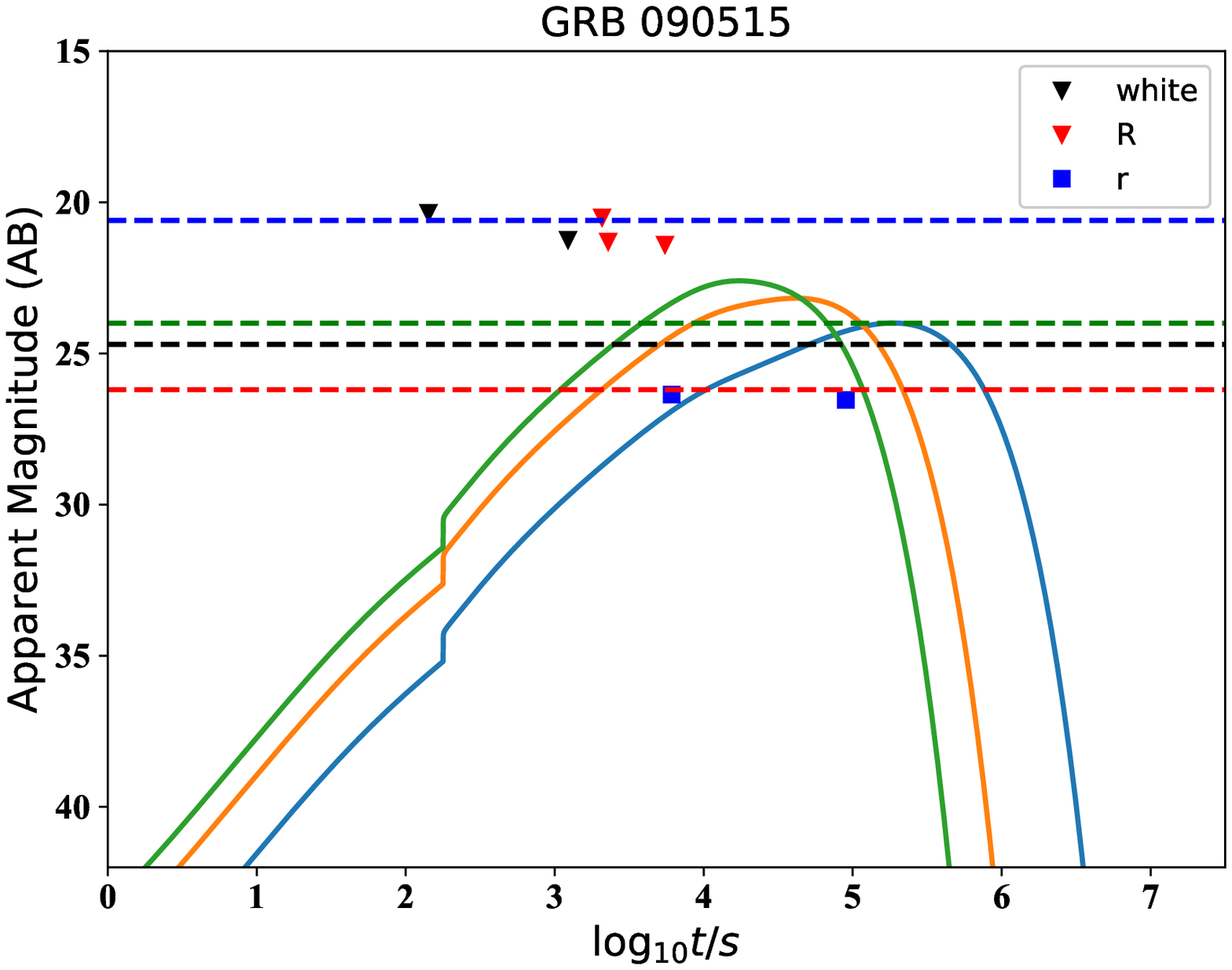}
\includegraphics[width=.25\textwidth]{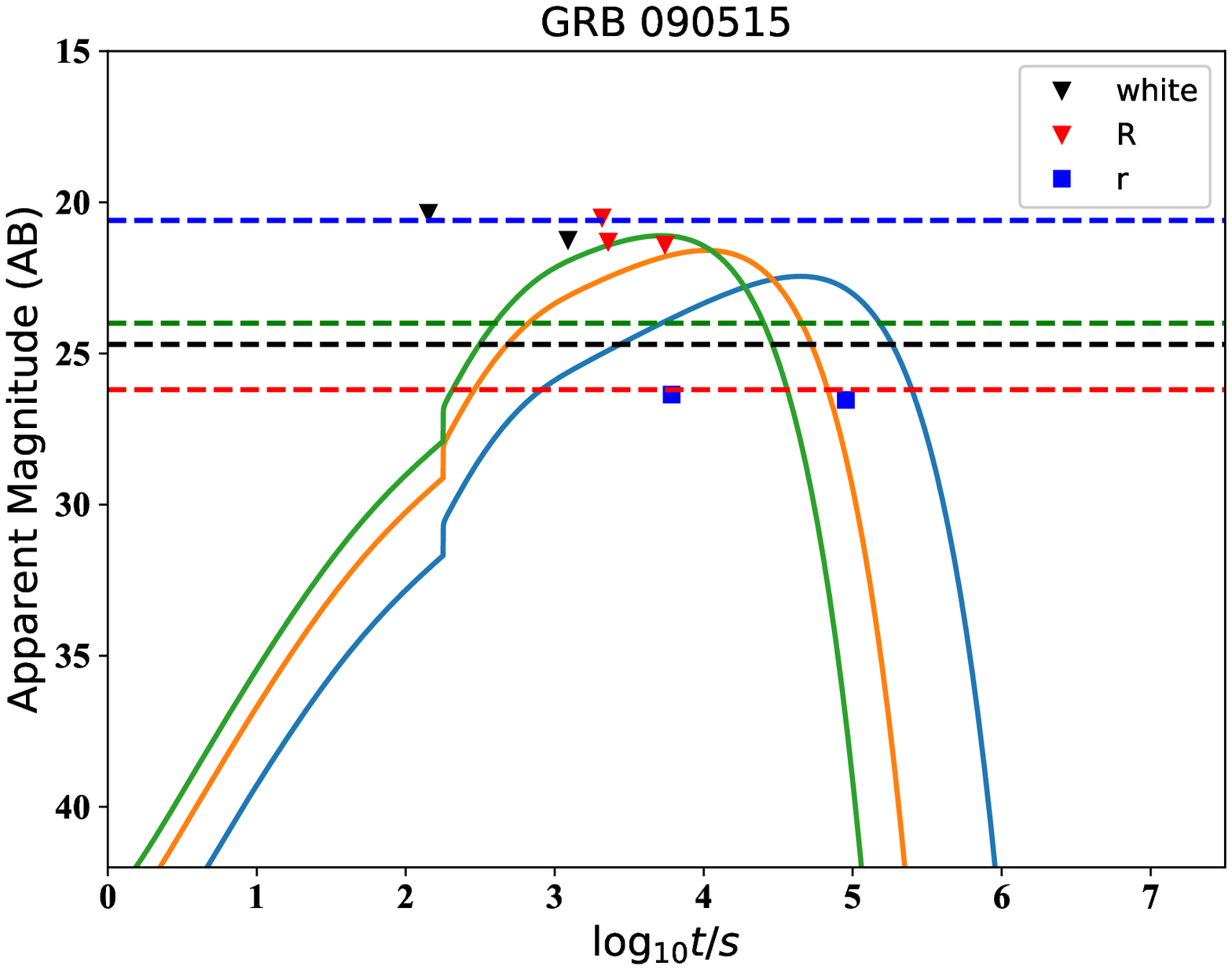}
\includegraphics[width=.25\textwidth]{GRB090515M001B01A01.eps}
\includegraphics[width=.25\textwidth]{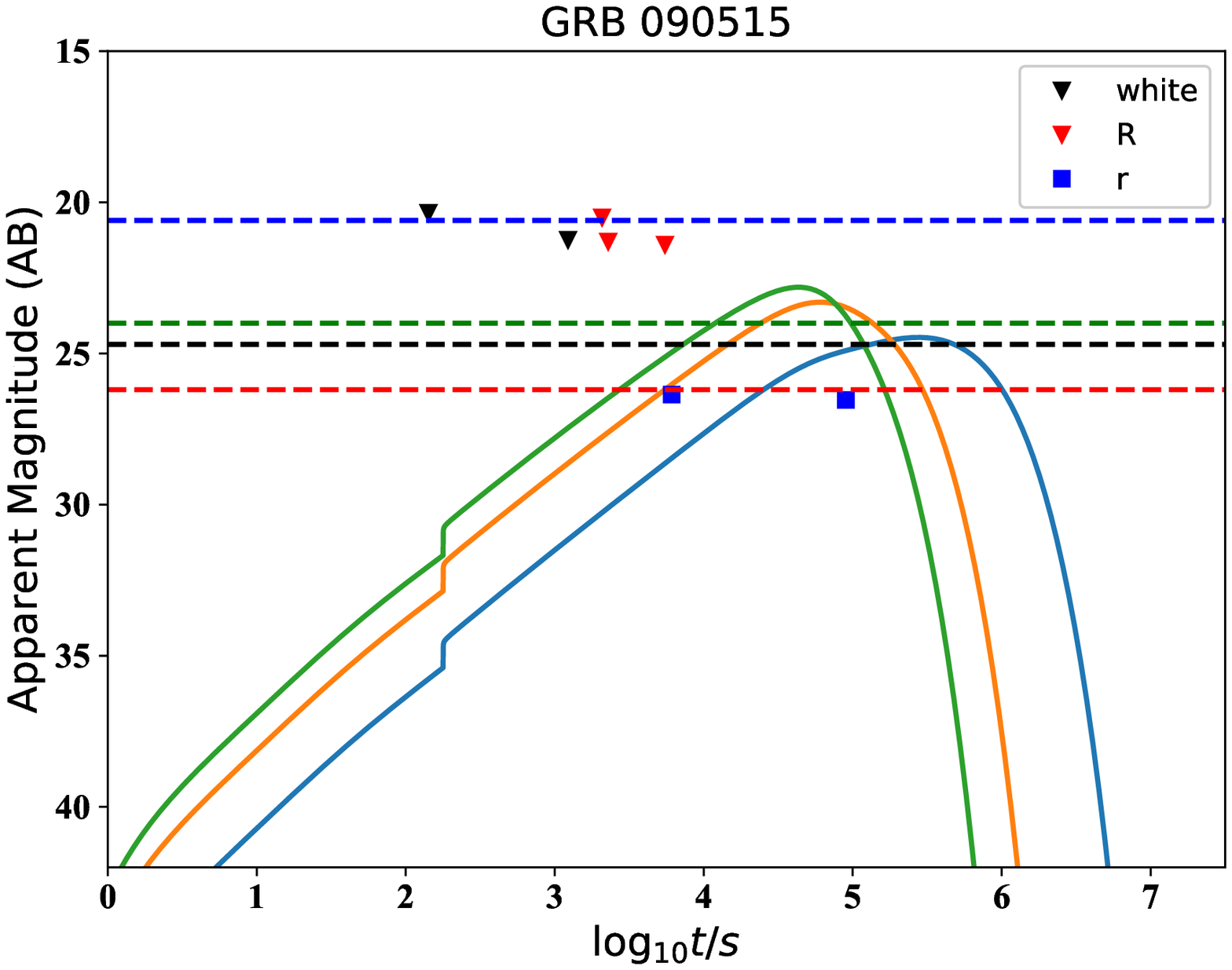}
\includegraphics[width=.25\textwidth]{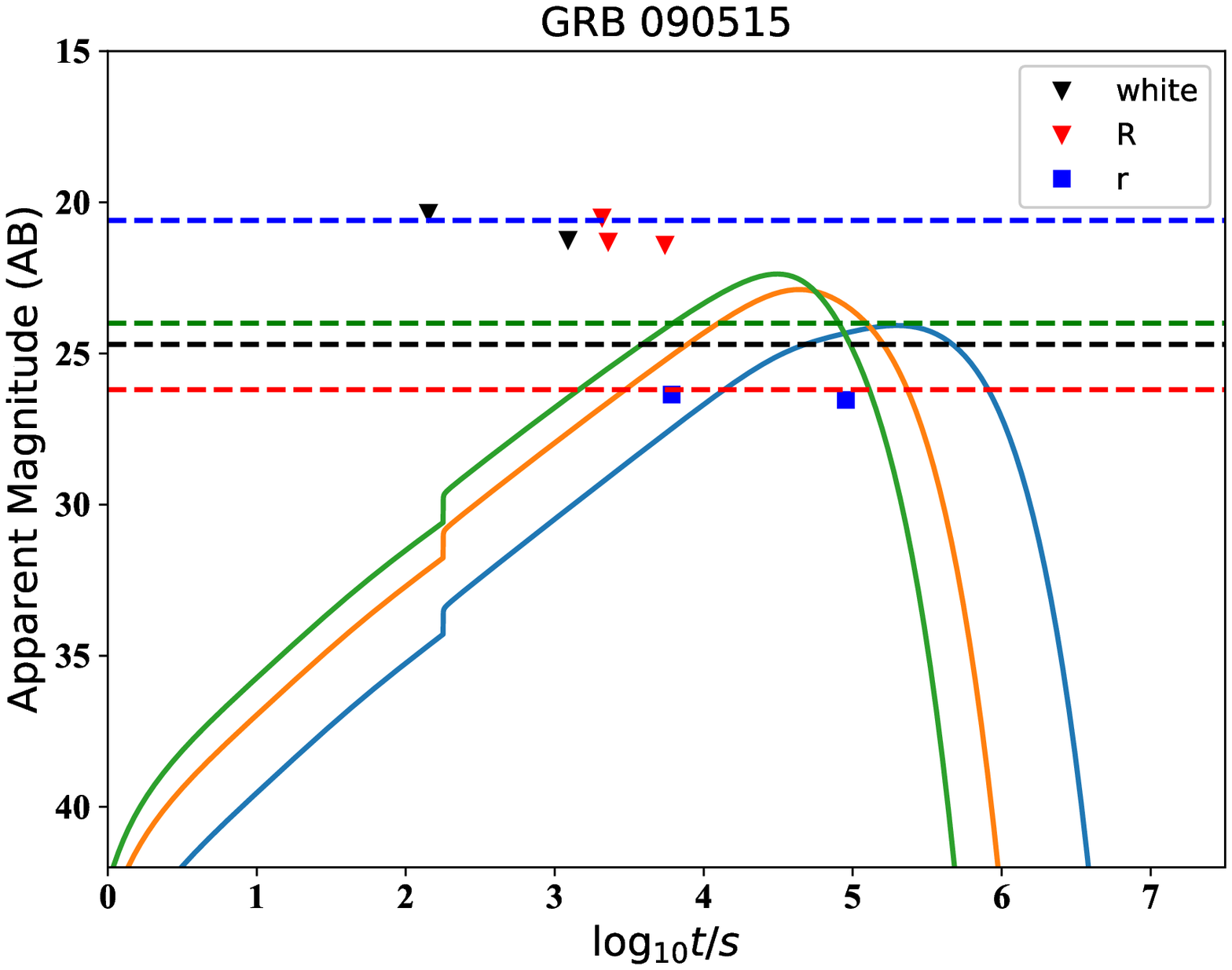}
\includegraphics[width=.25\textwidth]{GRB090515M001B01A01.eps}
\includegraphics[width=.25\textwidth]{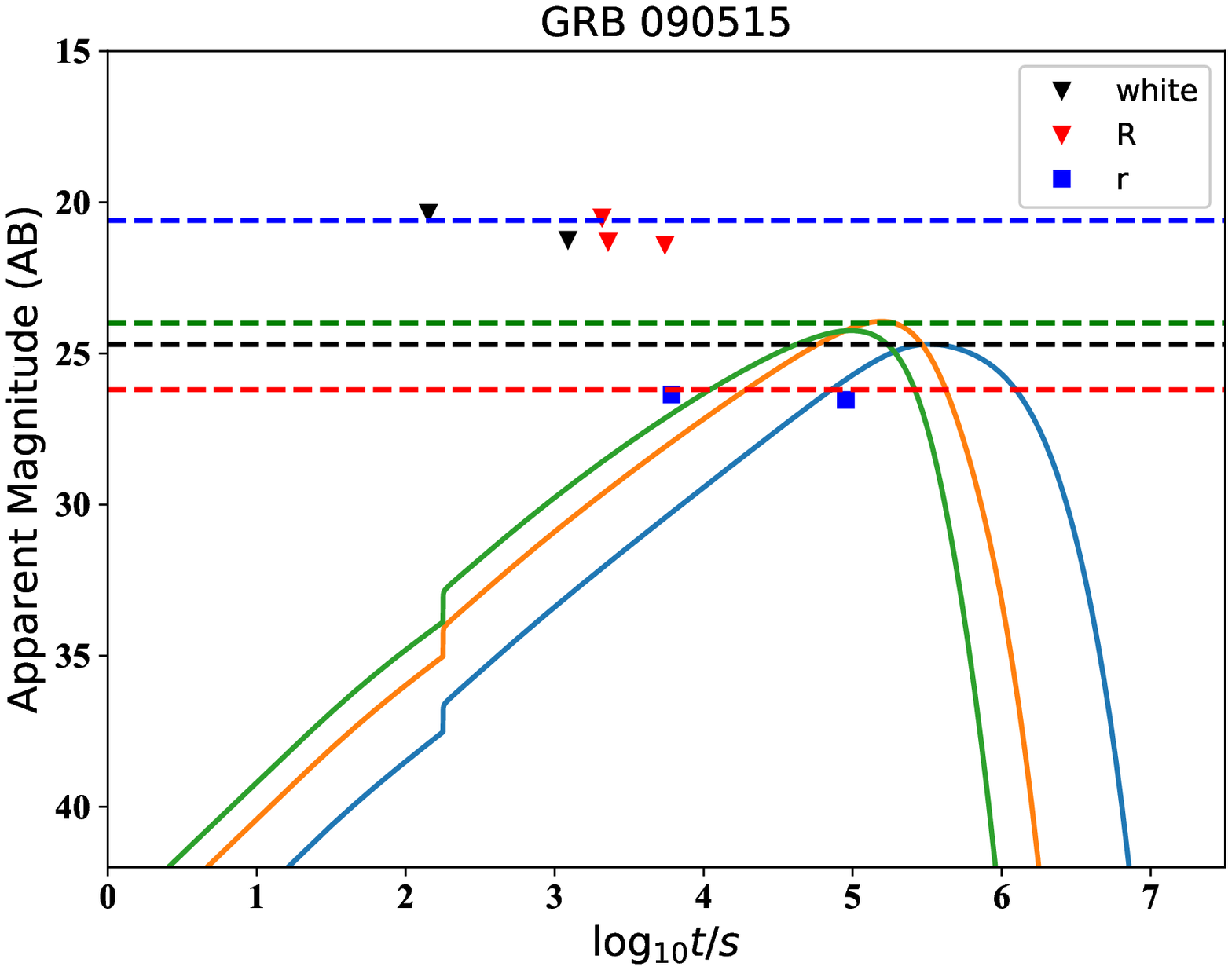}
\includegraphics[width=.25\textwidth]{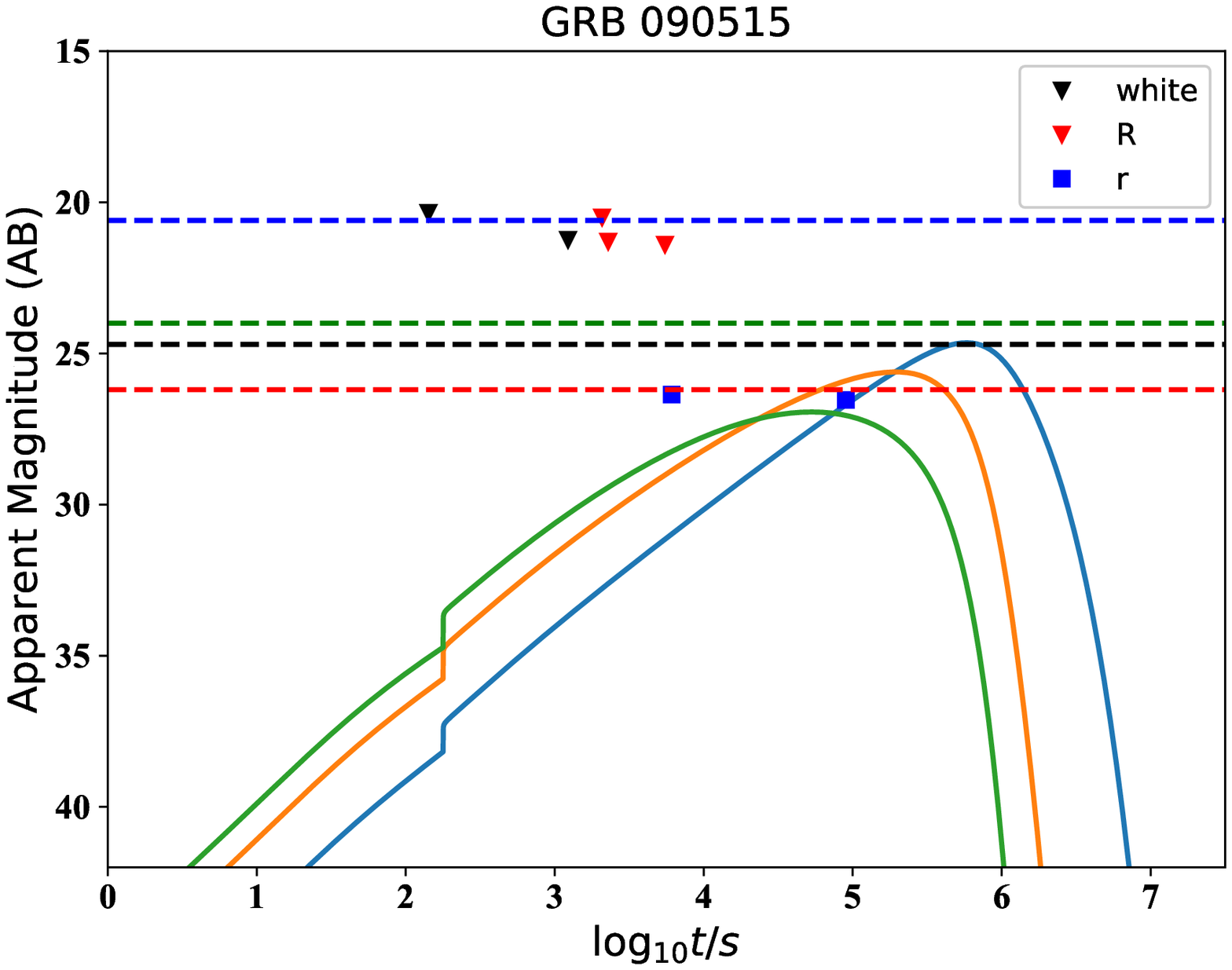}
\caption{Top, middle, and bottom panels are similar with Figure \ref{GRB060801}, but for GRB 090515.
The observed upper limit of optical data (triangles) and real data in r-band are
taken from Fong \& Berger (2013).}
\label{GRB090515}
\end{figure}

\begin{figure}
\centering
\includegraphics[width=.25\textwidth]{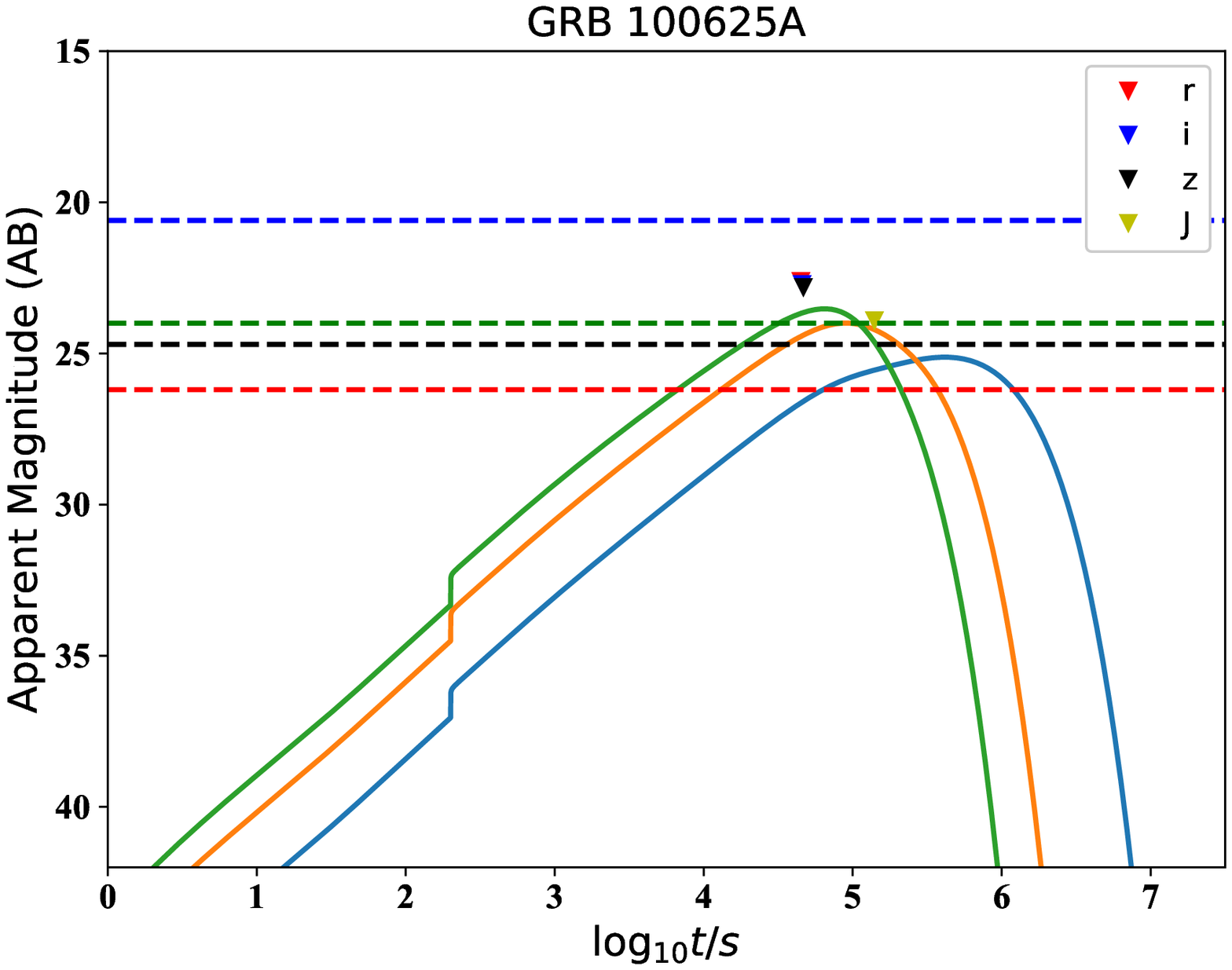}
\includegraphics[width=.25\textwidth]{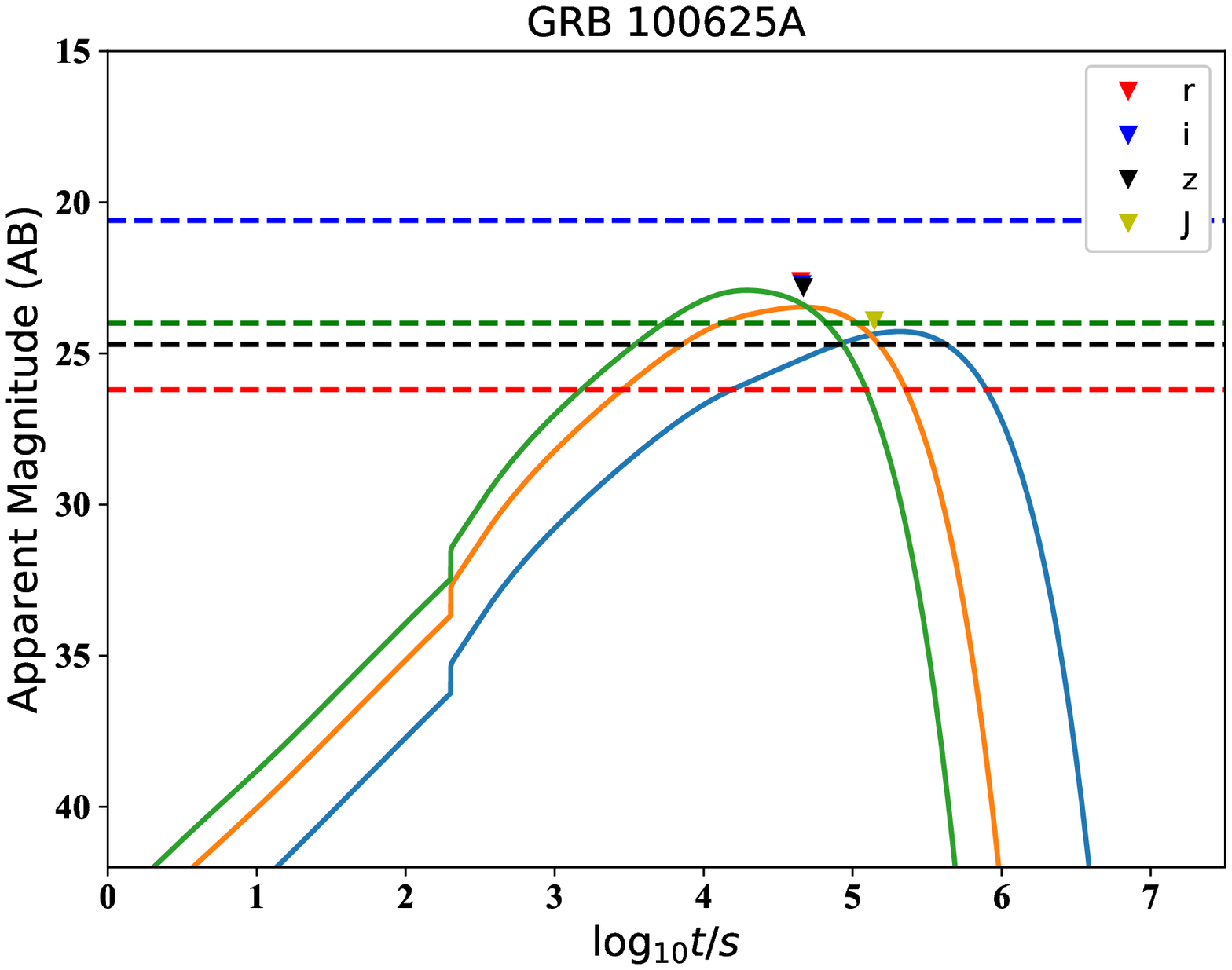}
\includegraphics[width=.25\textwidth]{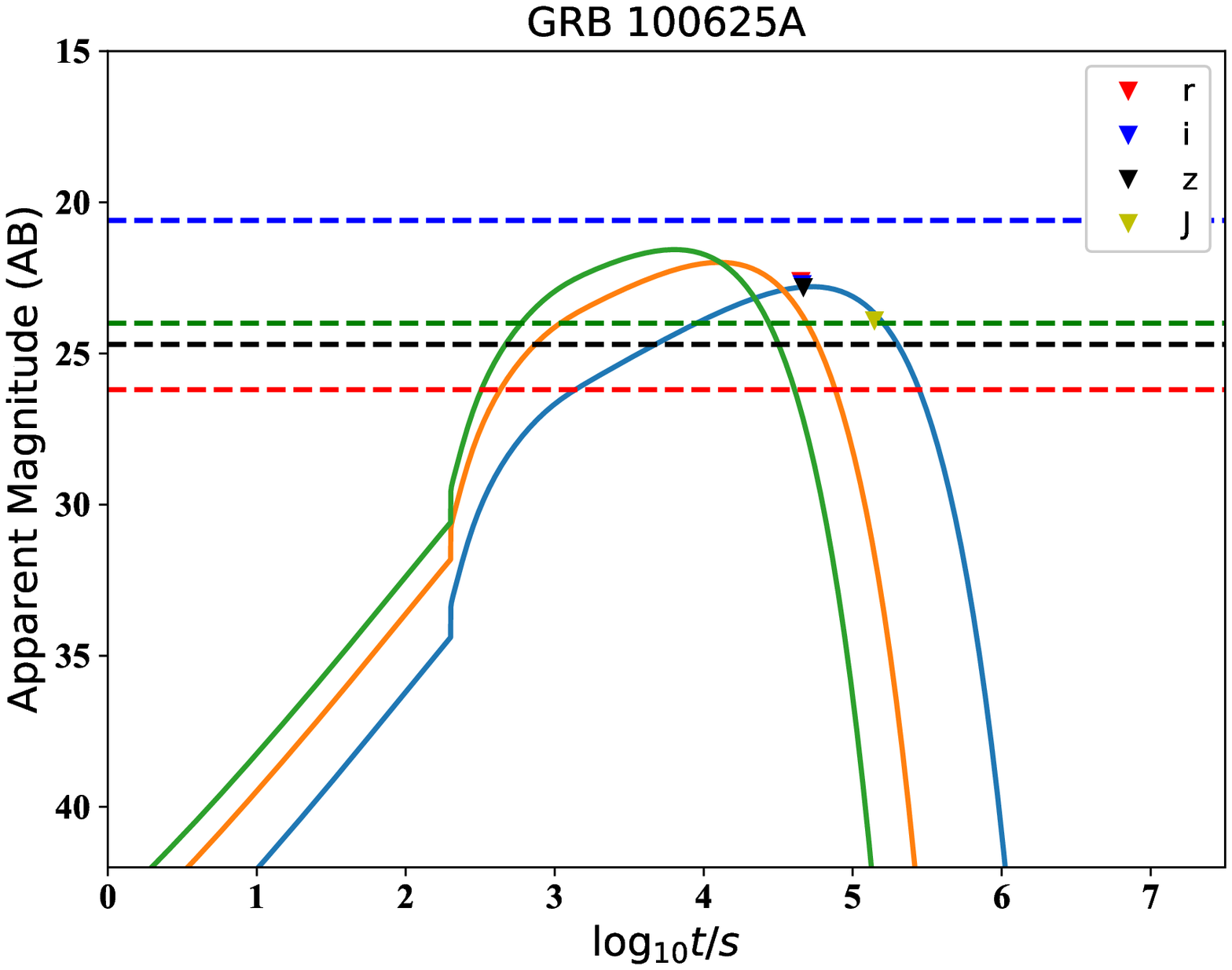}
\includegraphics[width=.25\textwidth]{GRB100625AM001B01A01.eps}
\includegraphics[width=.25\textwidth]{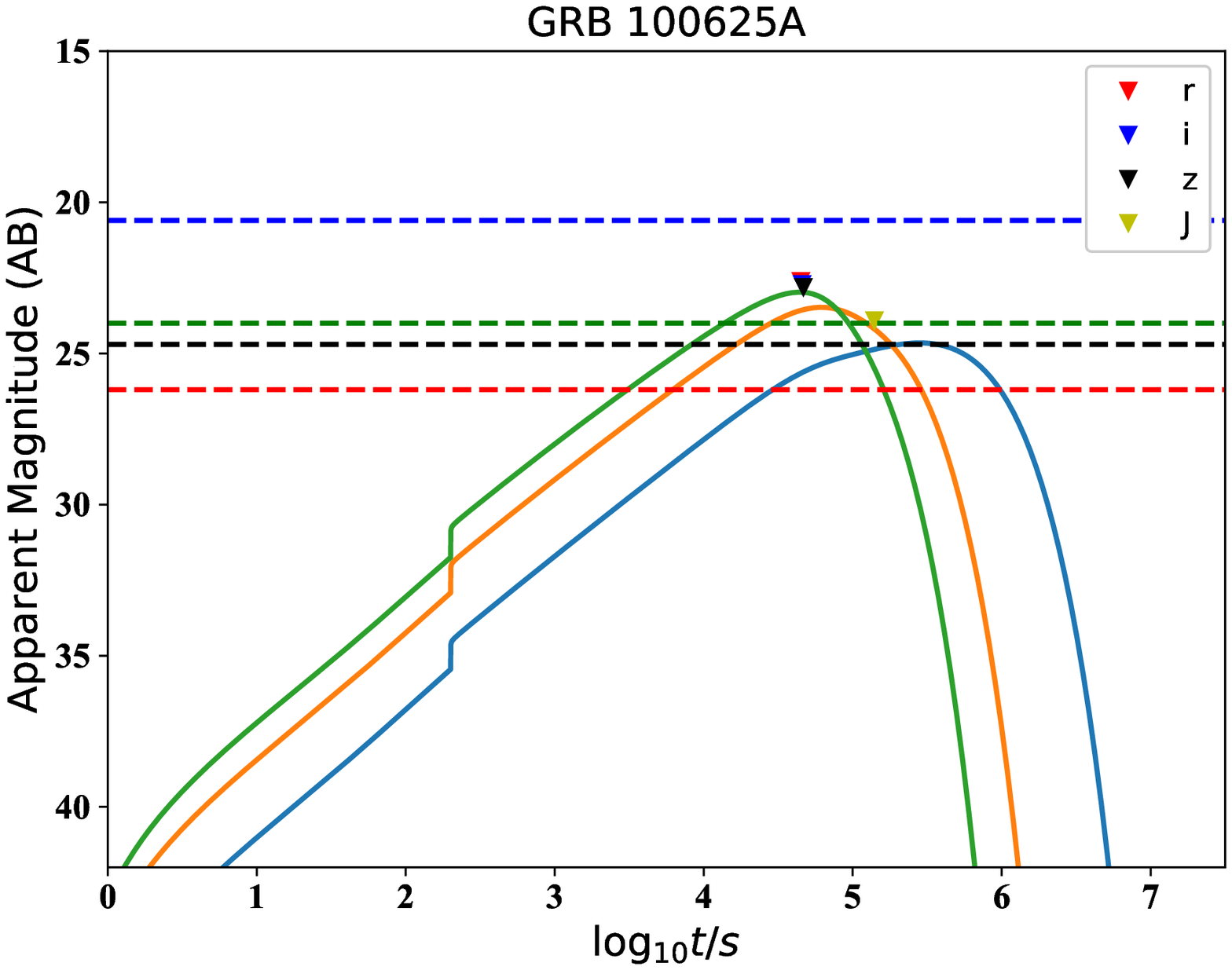}
\includegraphics[width=.25\textwidth]{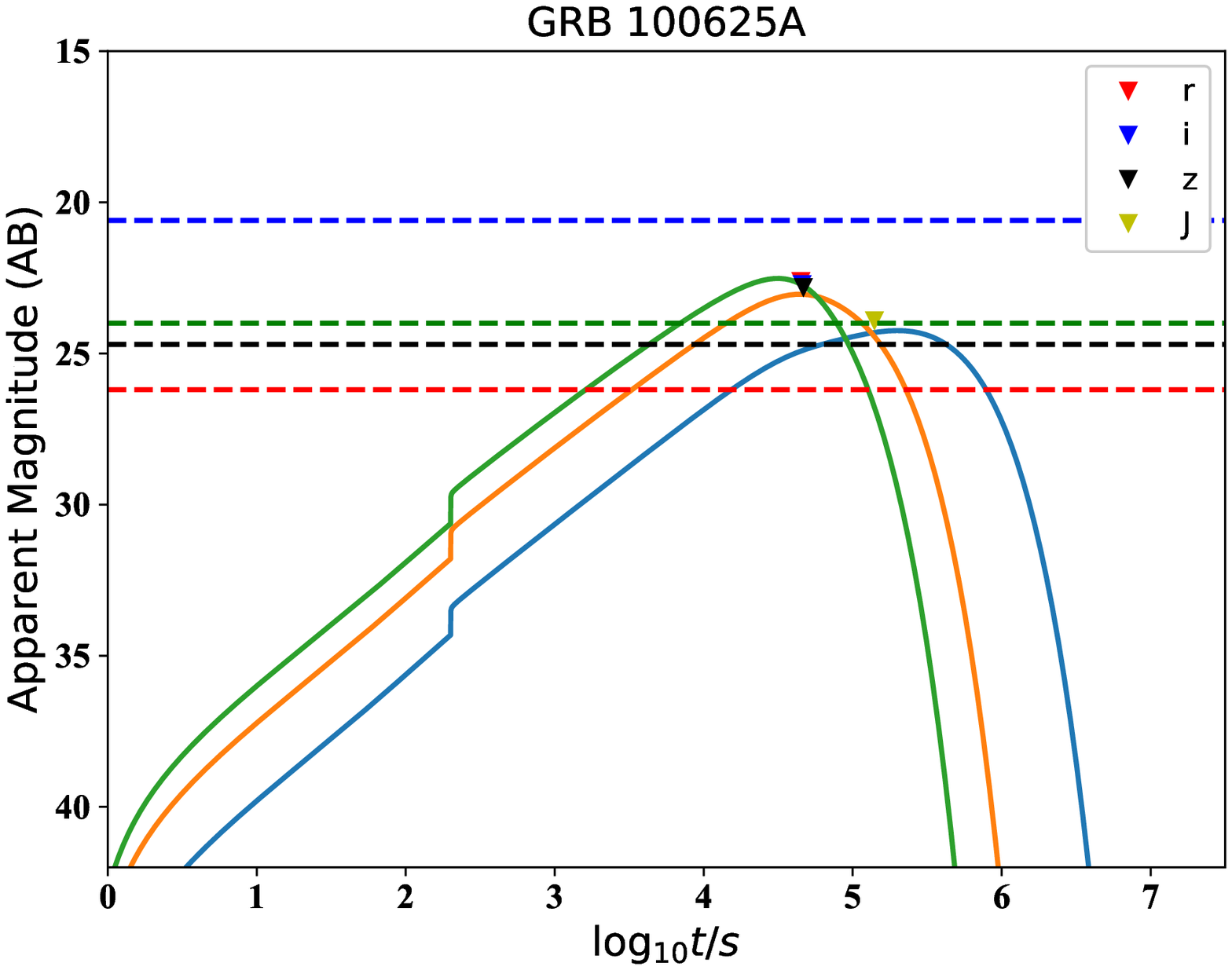}
\includegraphics[width=.25\textwidth]{GRB100625AM001B01A01.eps}
\includegraphics[width=.25\textwidth]{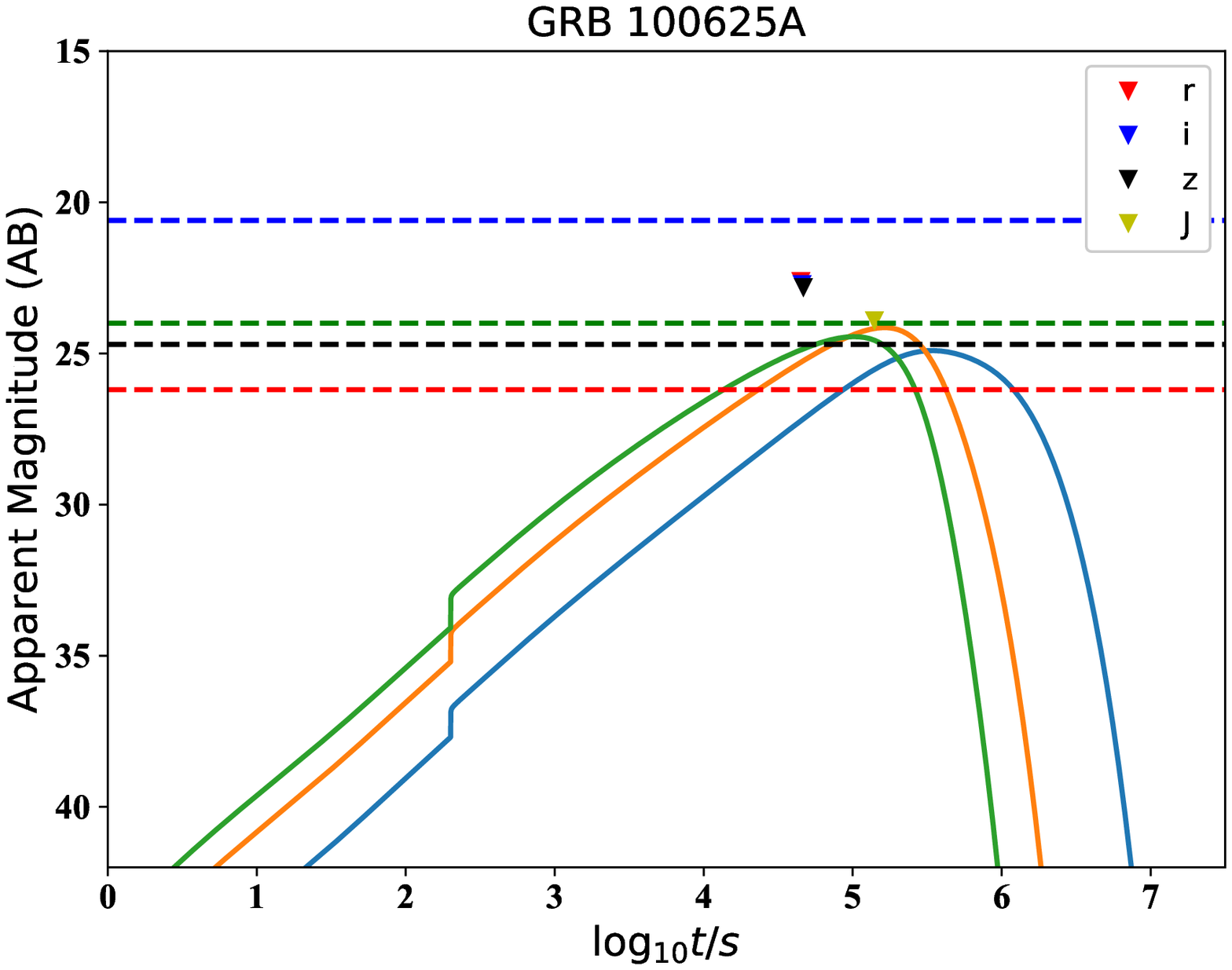}
\includegraphics[width=.25\textwidth]{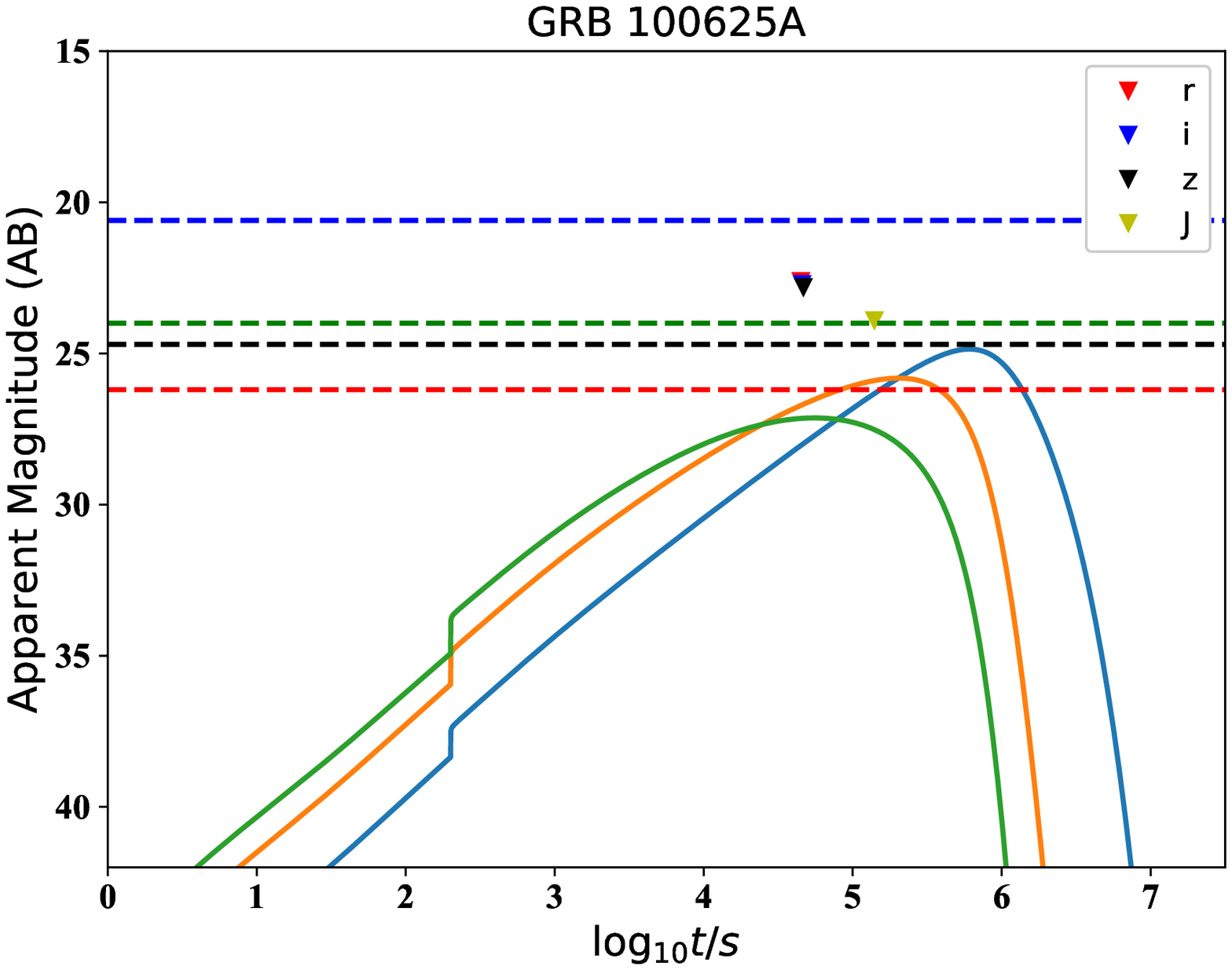}
\caption{Top, middle, and bottom panels are similar with Figure \ref{GRB060801}, but for GRB 100625A.
The observed upper limit of optical data (triangles) are
taken from Fong et al. (2013).}
\label{GRB100625A}
\end{figure}

\begin{figure}
\centering
\includegraphics[width=.25\textwidth]{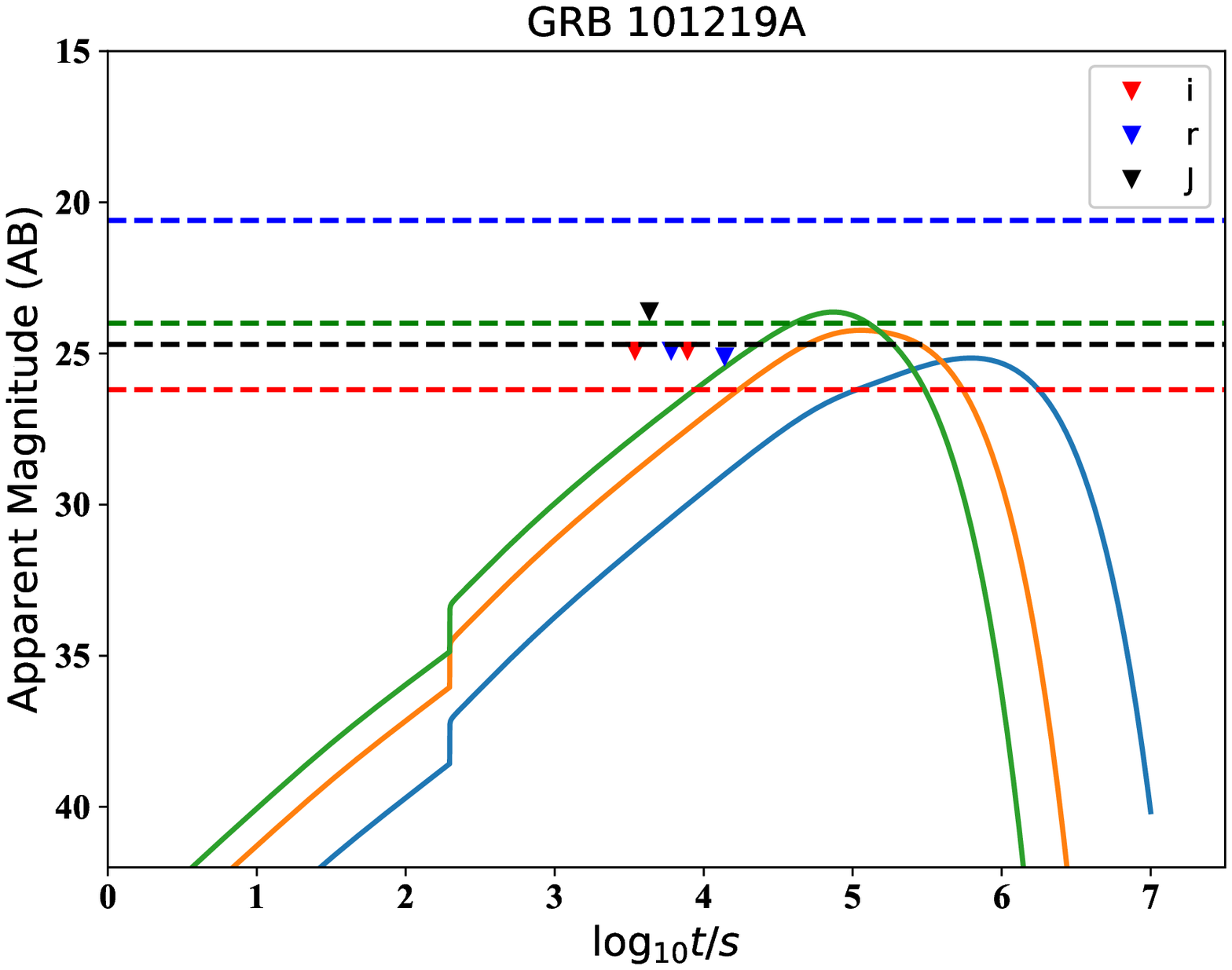}
\includegraphics[width=.25\textwidth]{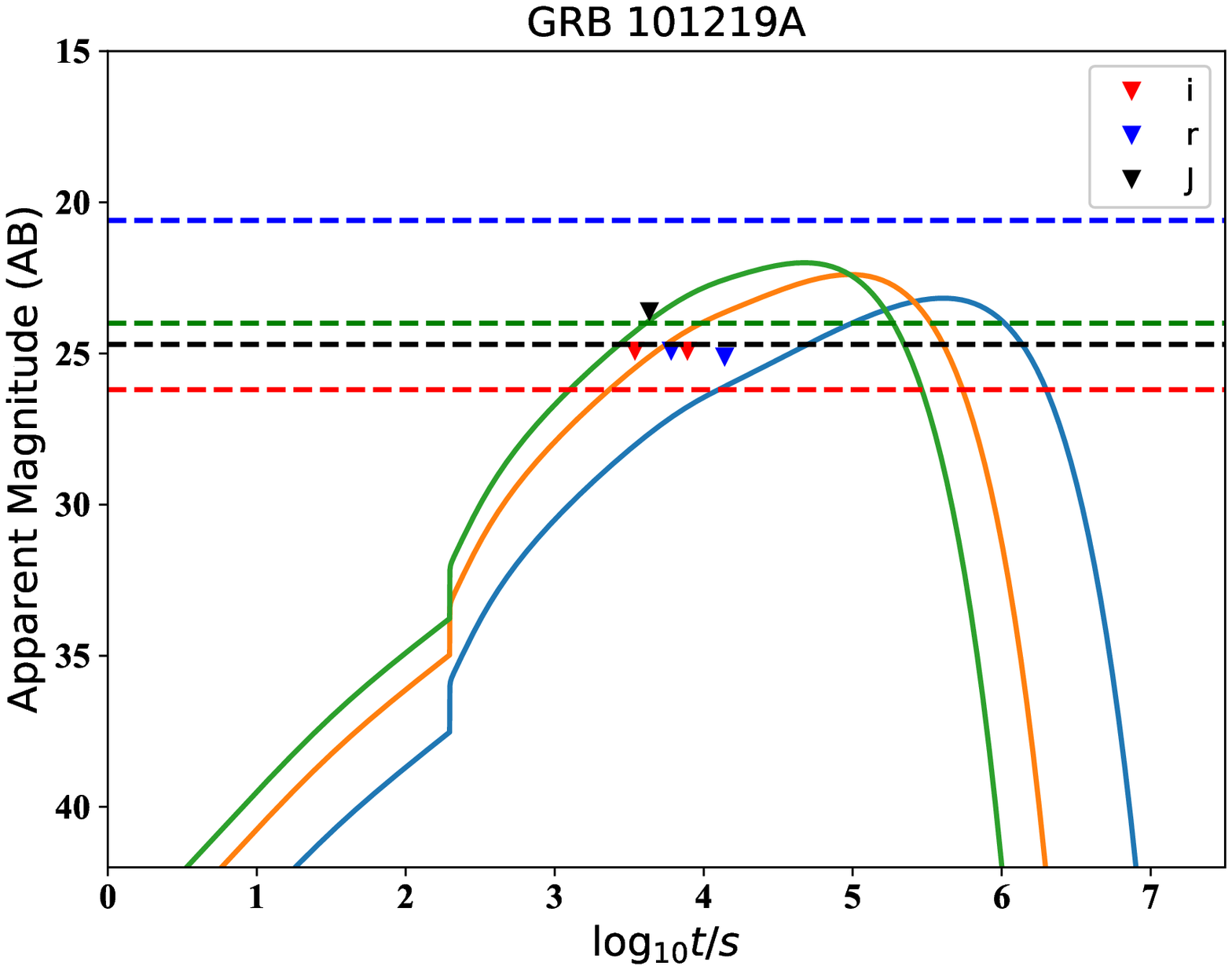}
\includegraphics[width=.25\textwidth]{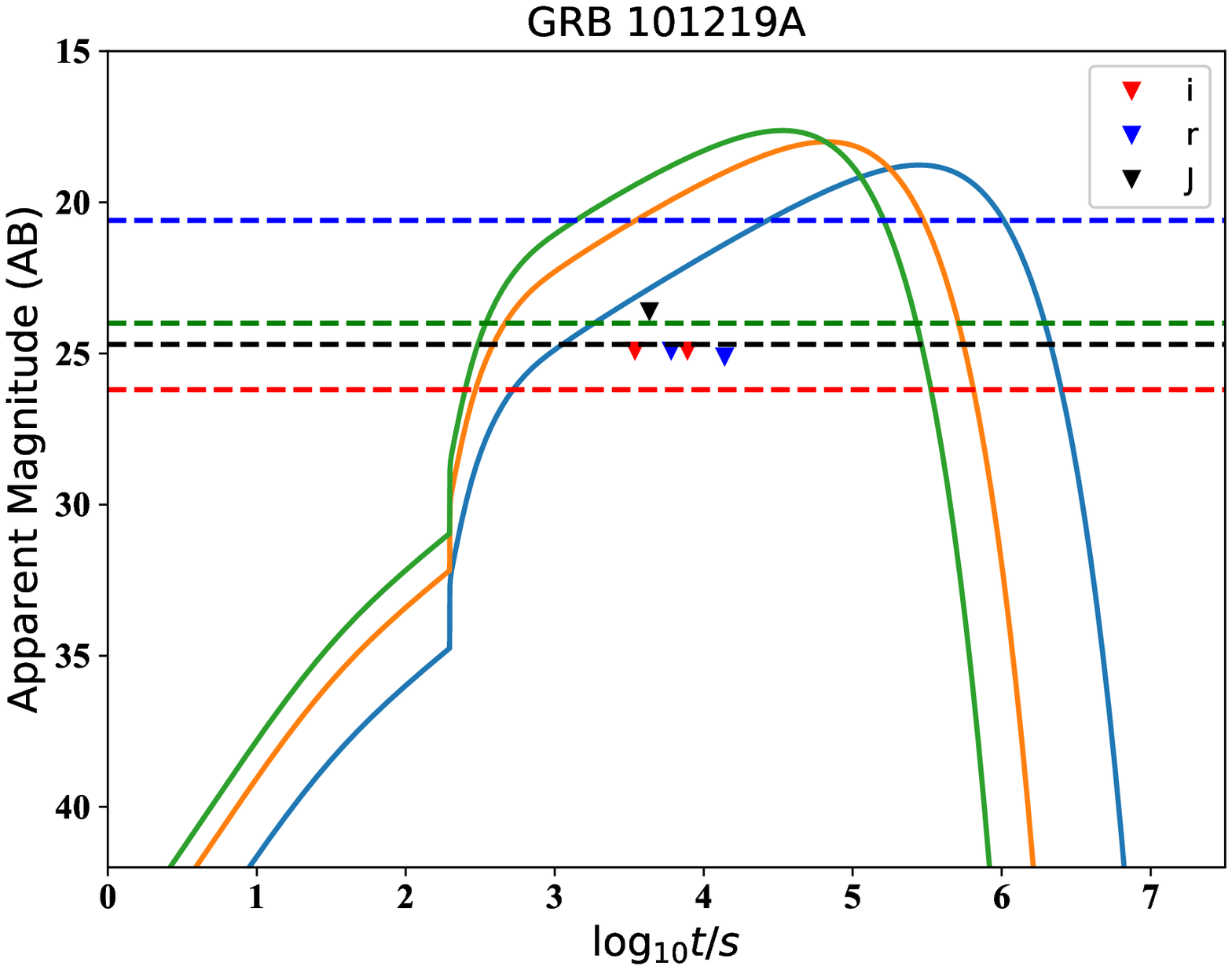}
\includegraphics[width=.25\textwidth]{GRB101219AM001B01A01.eps}
\includegraphics[width=.25\textwidth]{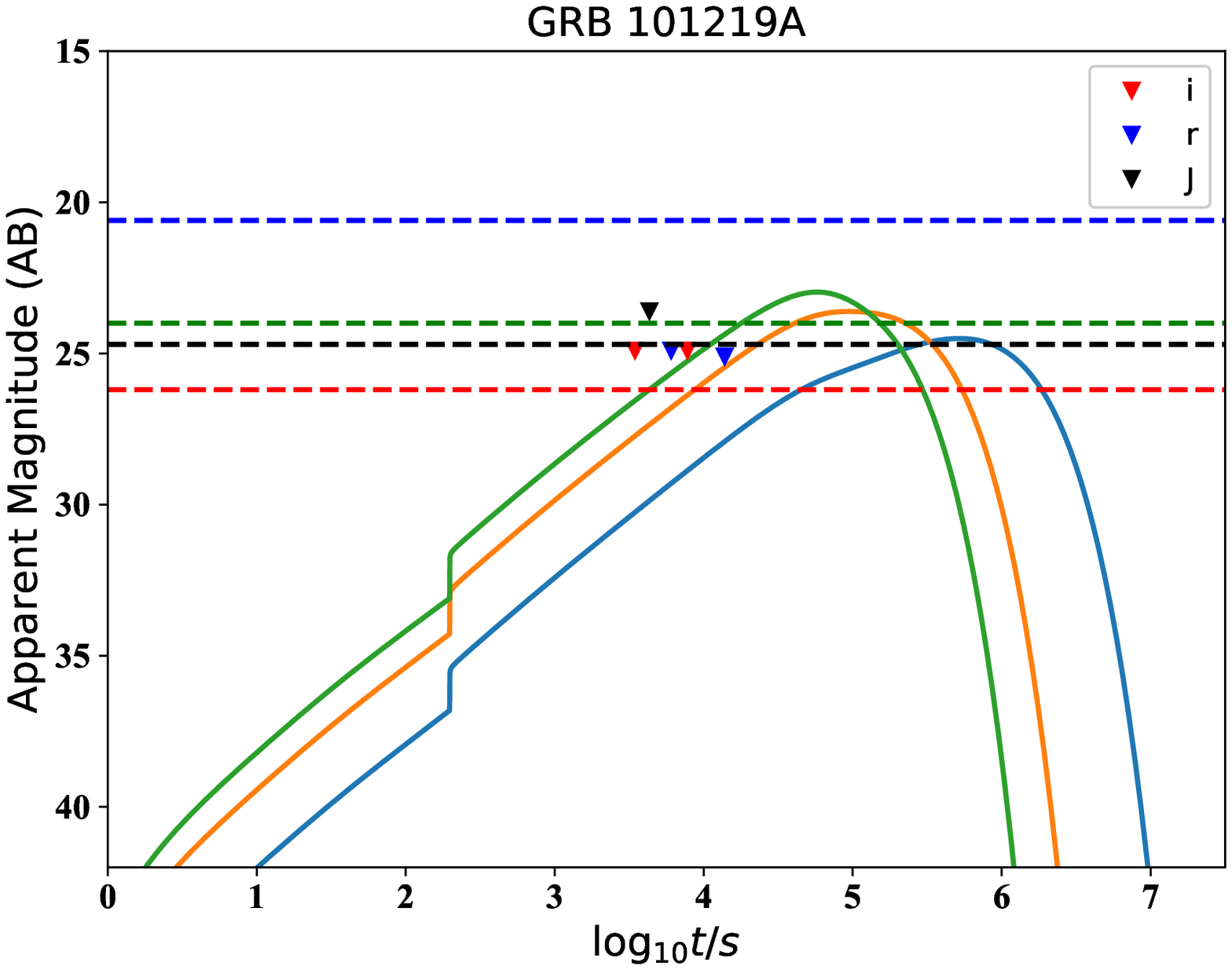}
\includegraphics[width=.25\textwidth]{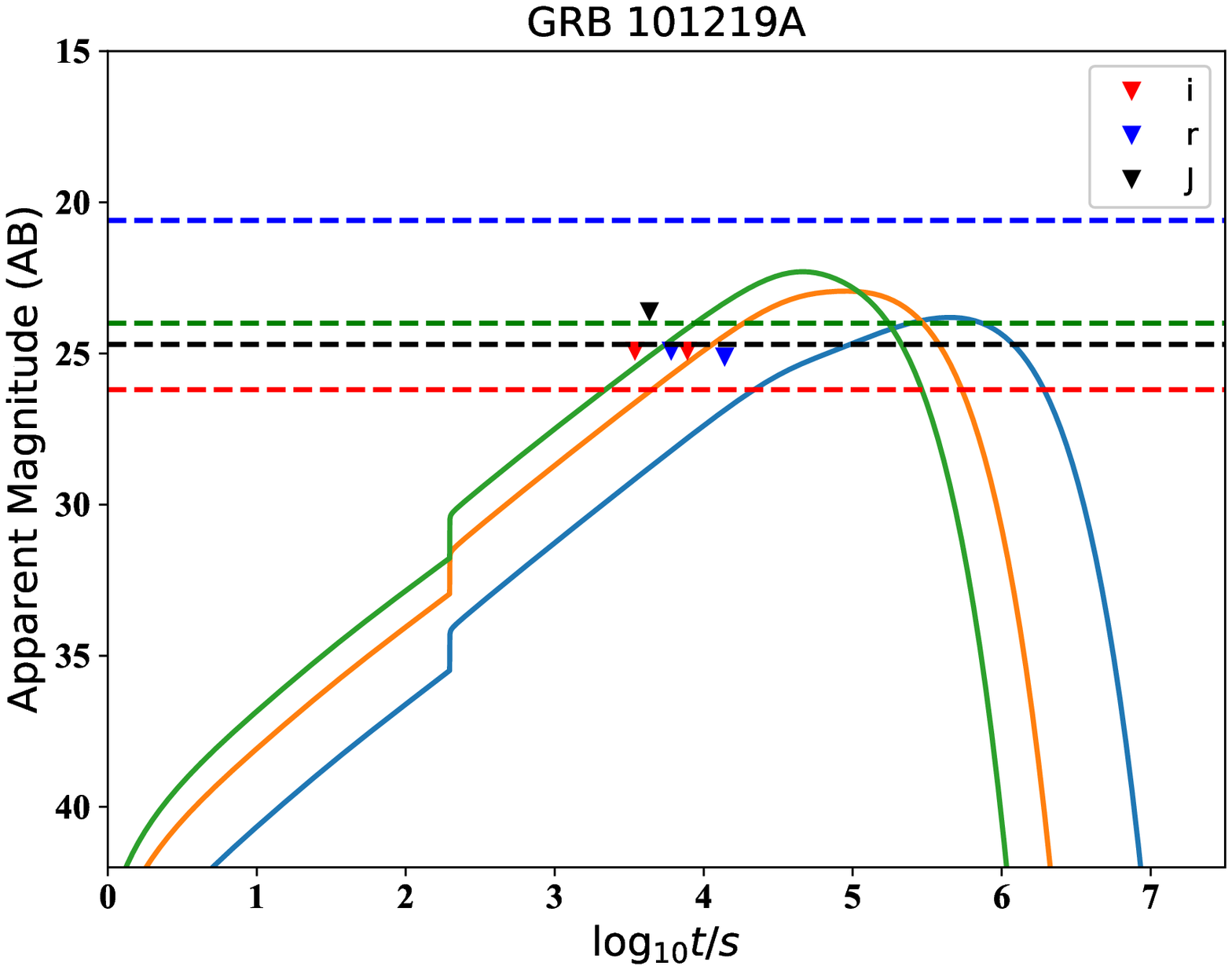}
\includegraphics[width=.25\textwidth]{GRB101219AM001B01A01.eps}
\includegraphics[width=.25\textwidth]{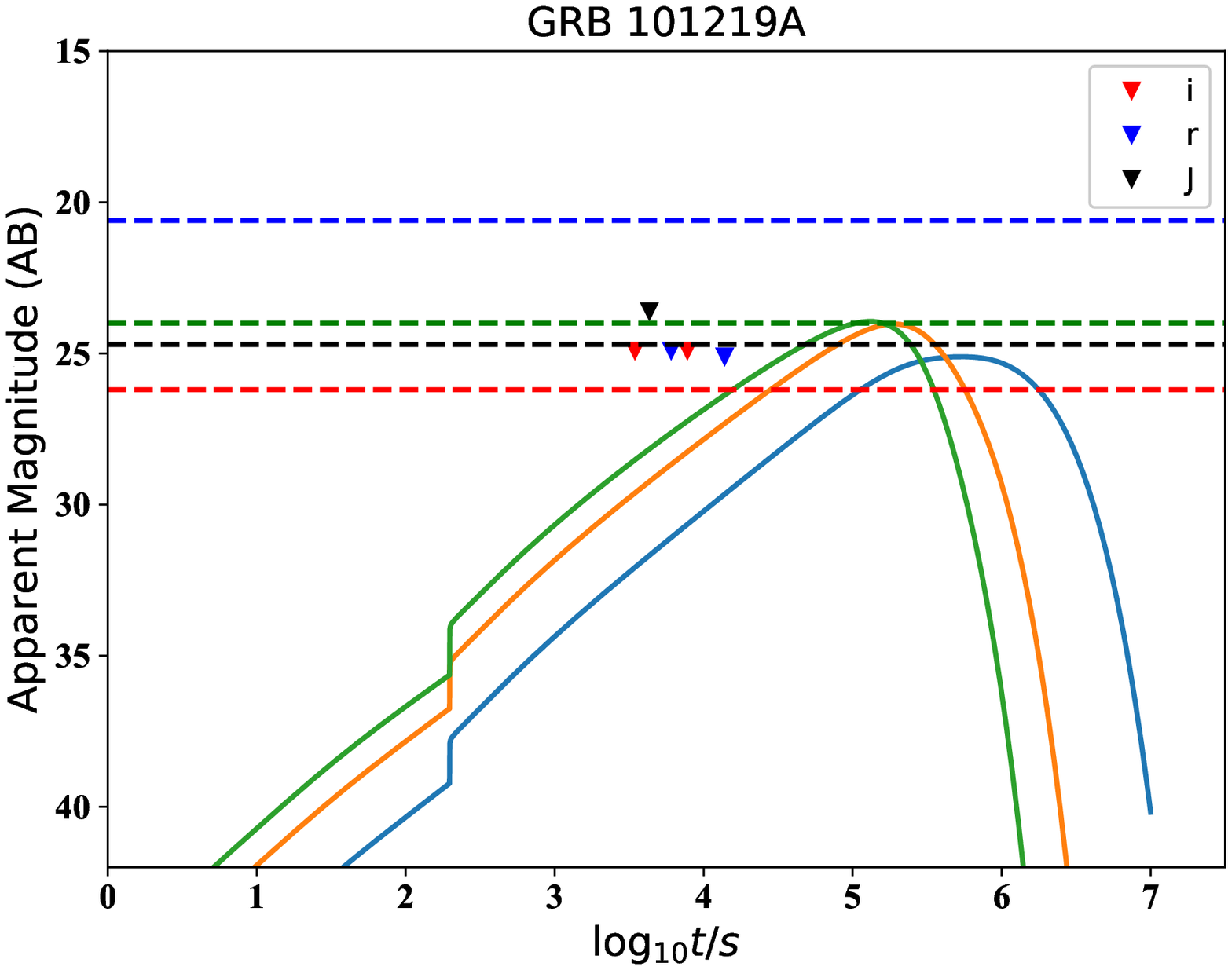}
\includegraphics[width=.25\textwidth]{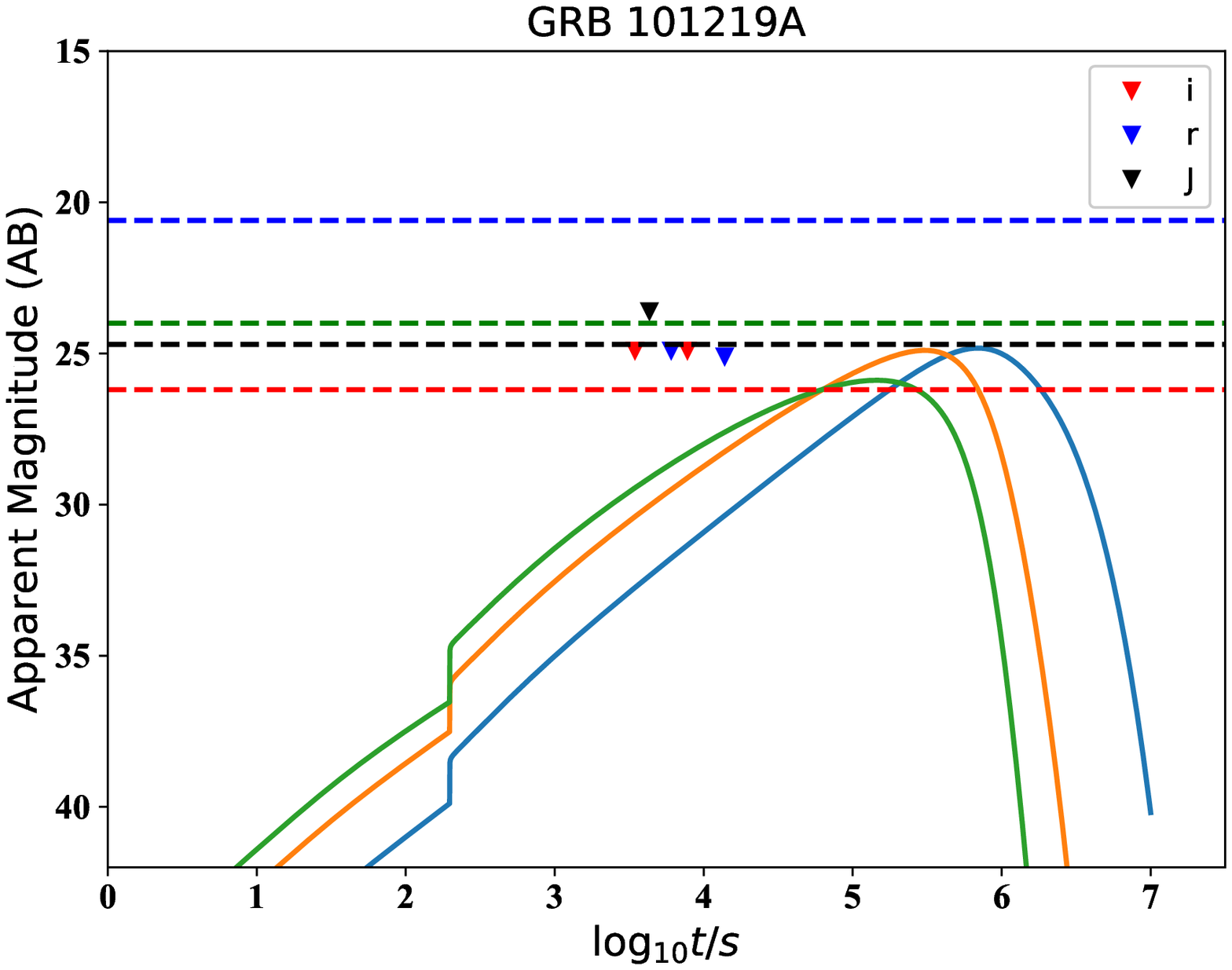}
\caption{Top, middle, and bottom panels are similar with Figure \ref{GRB060801}, but for GRB 101219A.
The observed upper limit of optical data (triangles) are
taken from Fong et al. (2013).}
\label{GRB101219A}
\end{figure}

\begin{figure}
\centering
\includegraphics[width=.25\textwidth]{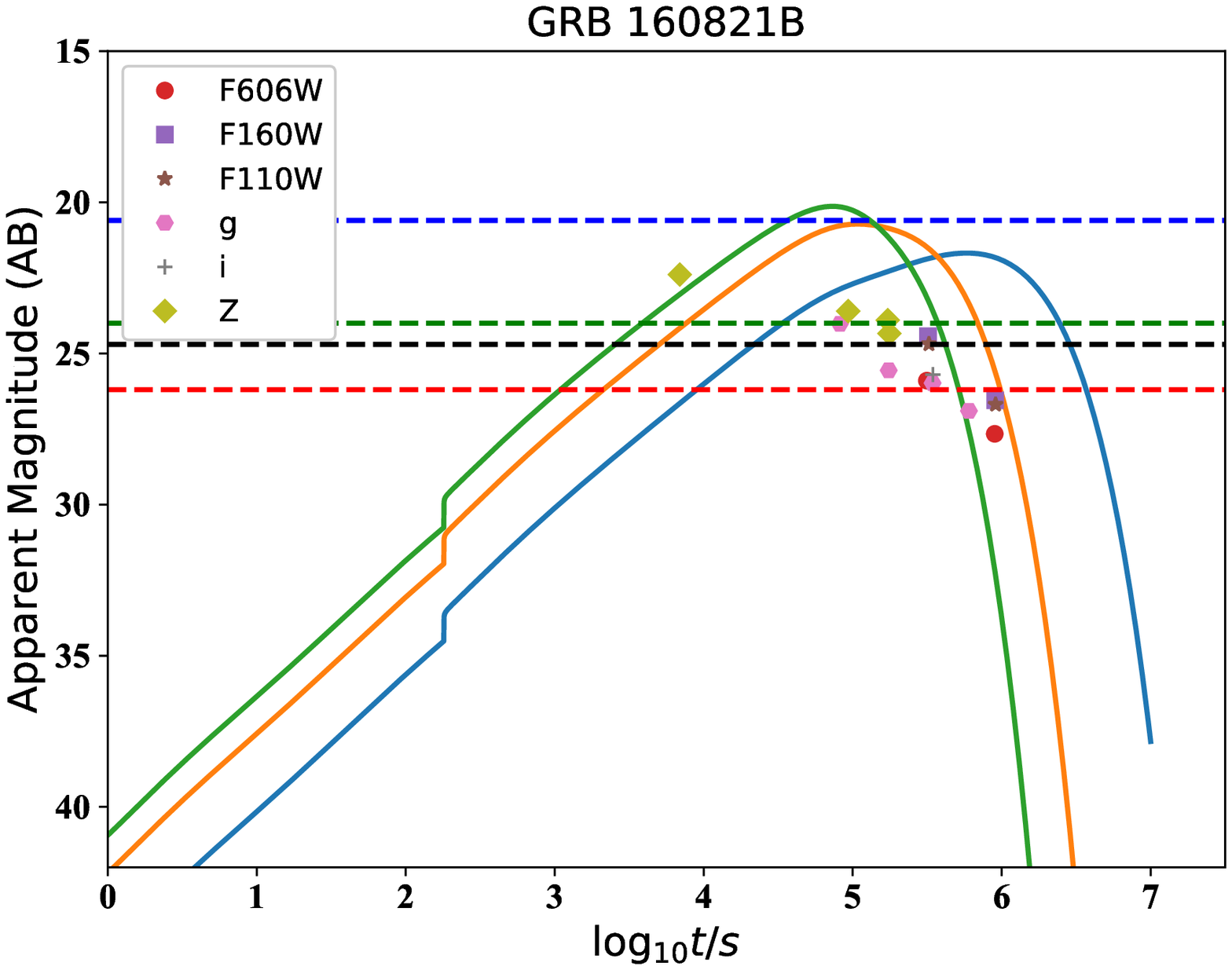}
\includegraphics[width=.25\textwidth]{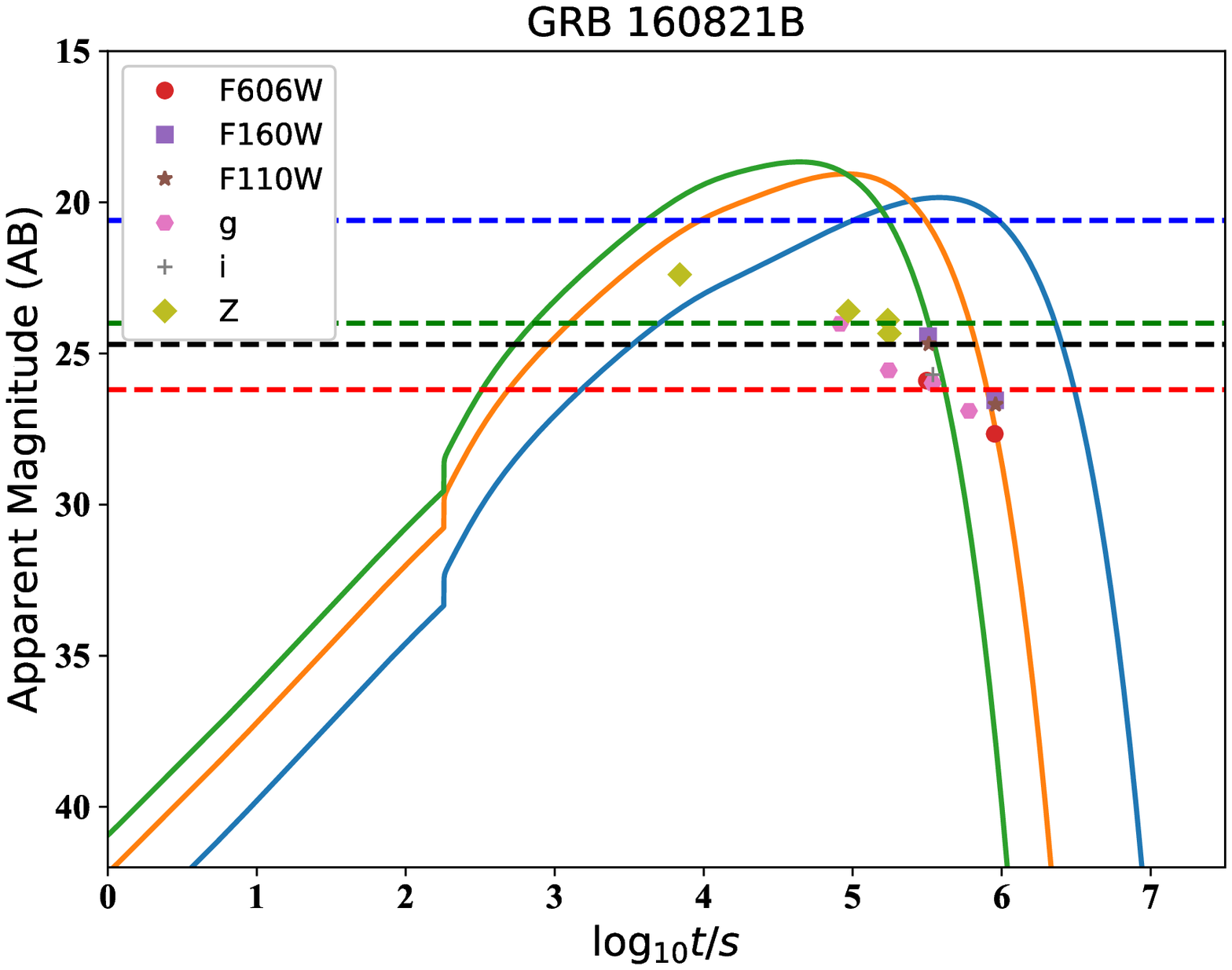}
\includegraphics[width=.25\textwidth]{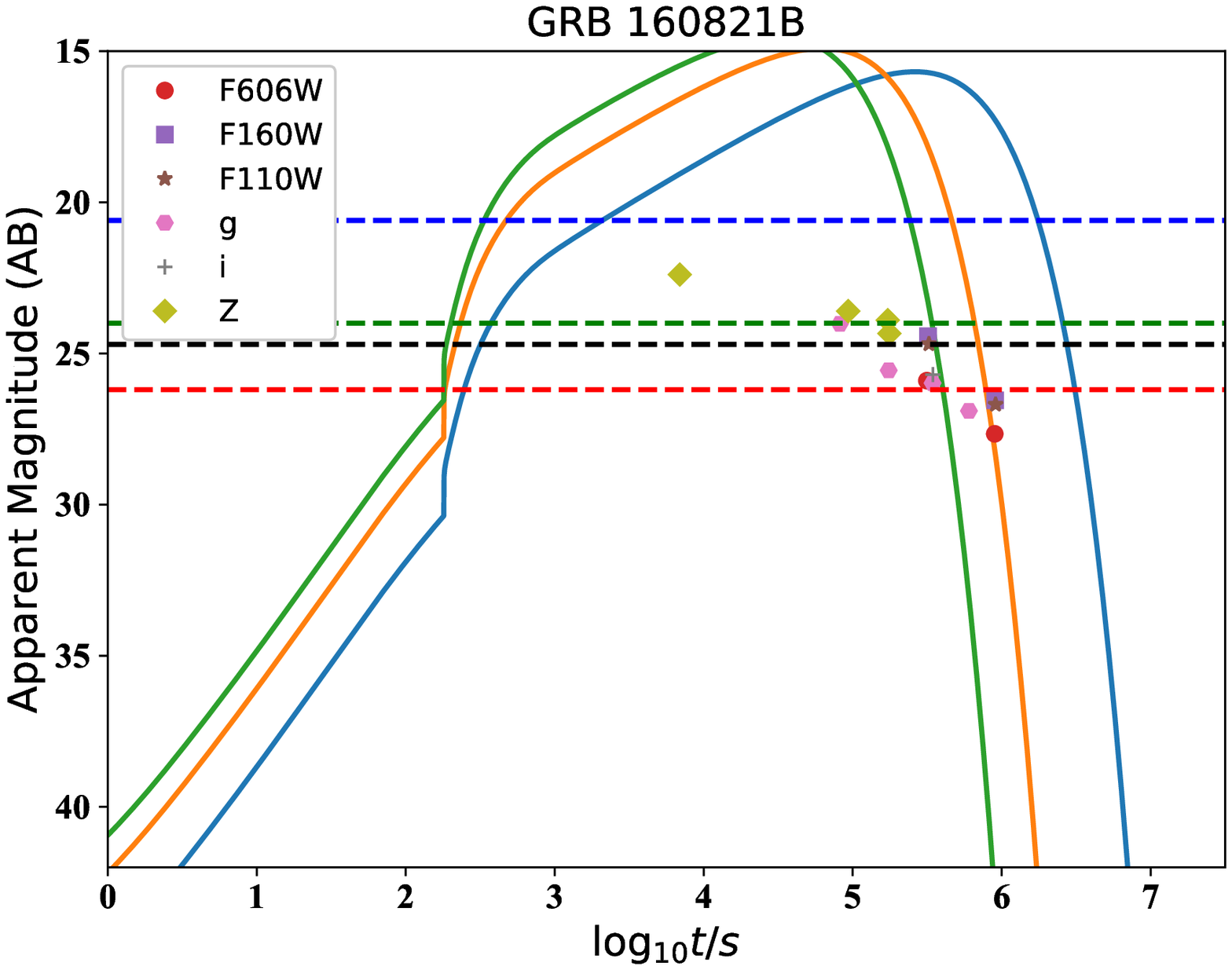}
\includegraphics[width=.25\textwidth]{GRB160821BM001B01A01.eps}
\includegraphics[width=.25\textwidth]{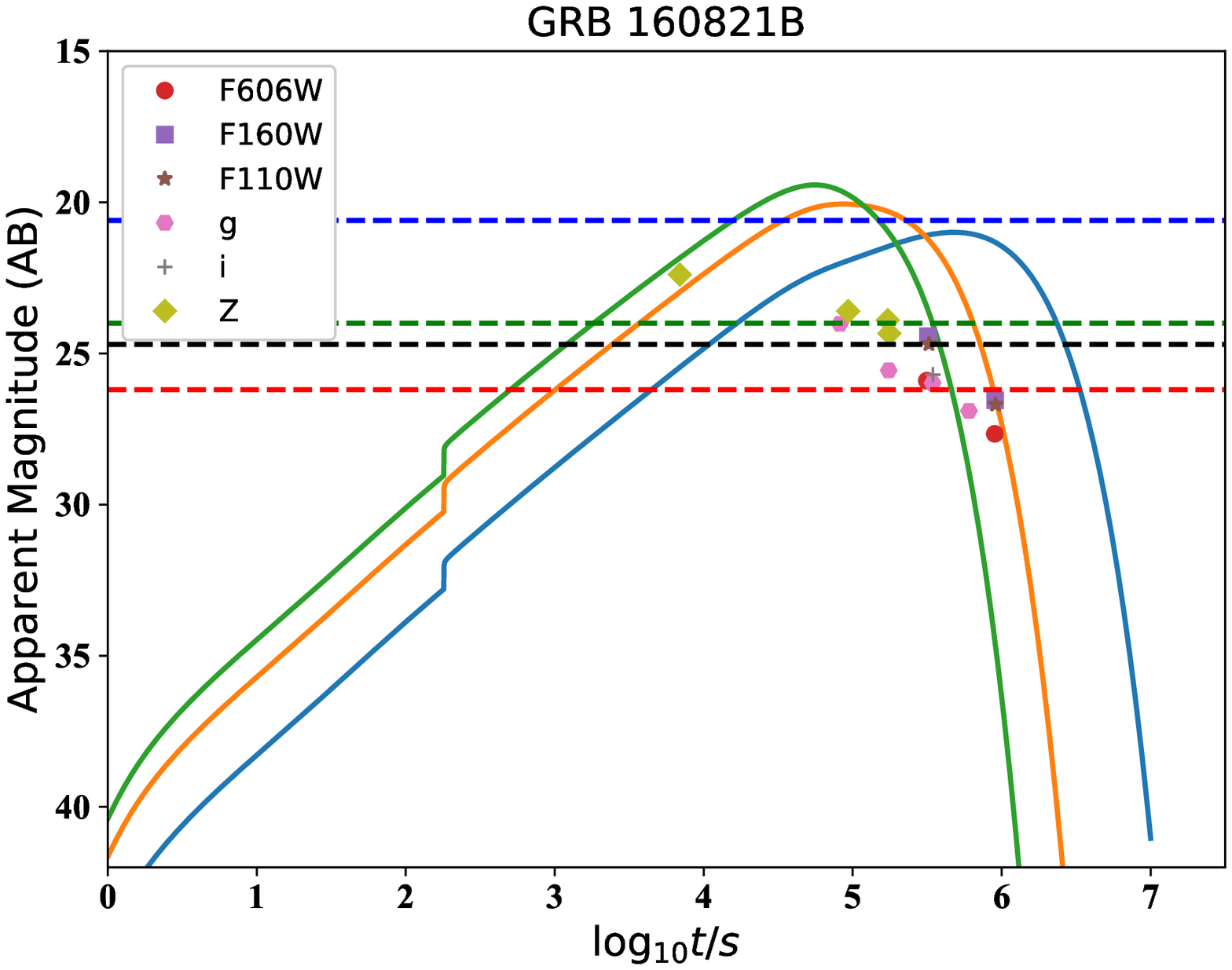}
\includegraphics[width=.25\textwidth]{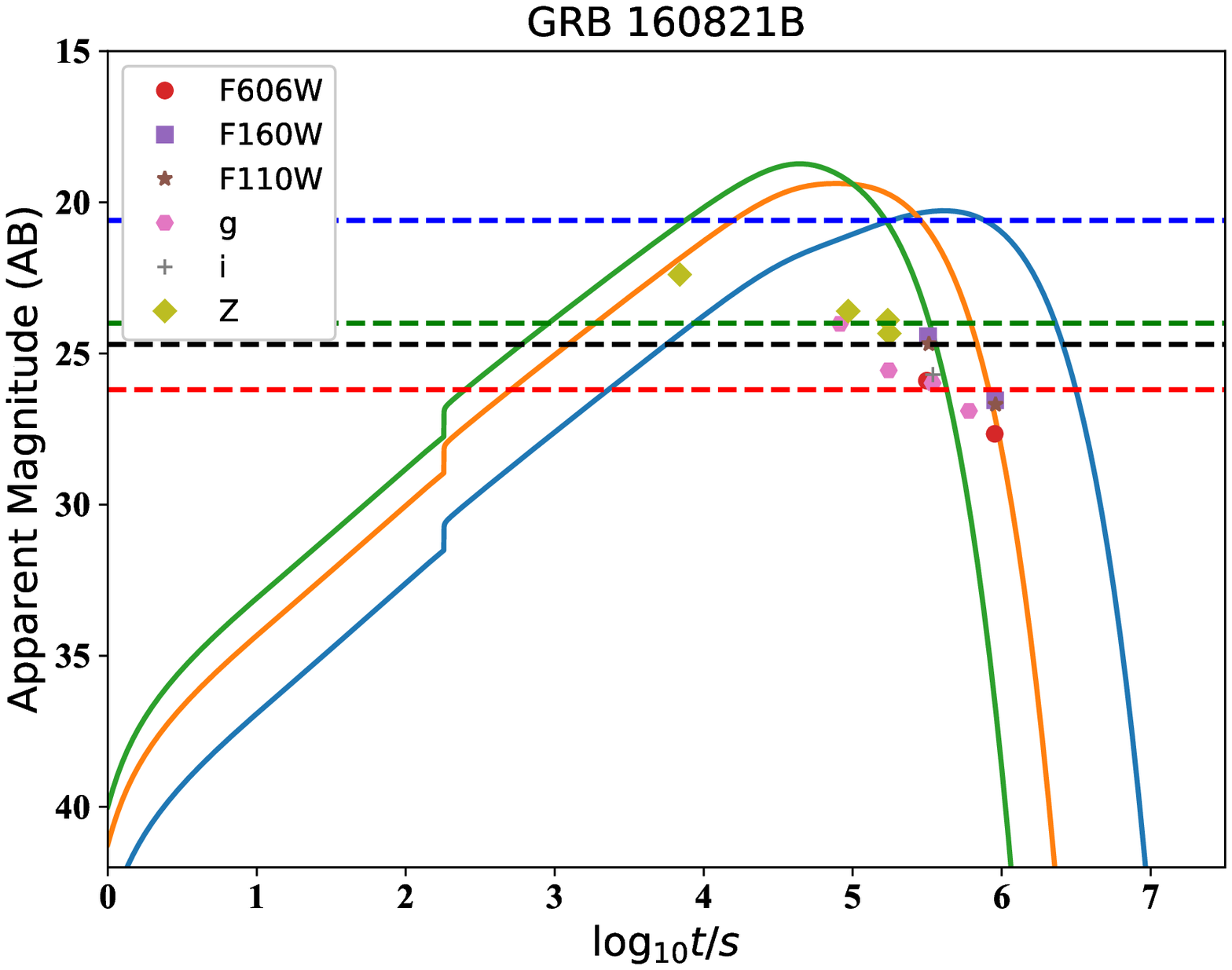}
\includegraphics[width=.25\textwidth]{GRB160821BM001B01A01.eps}
\includegraphics[width=.25\textwidth]{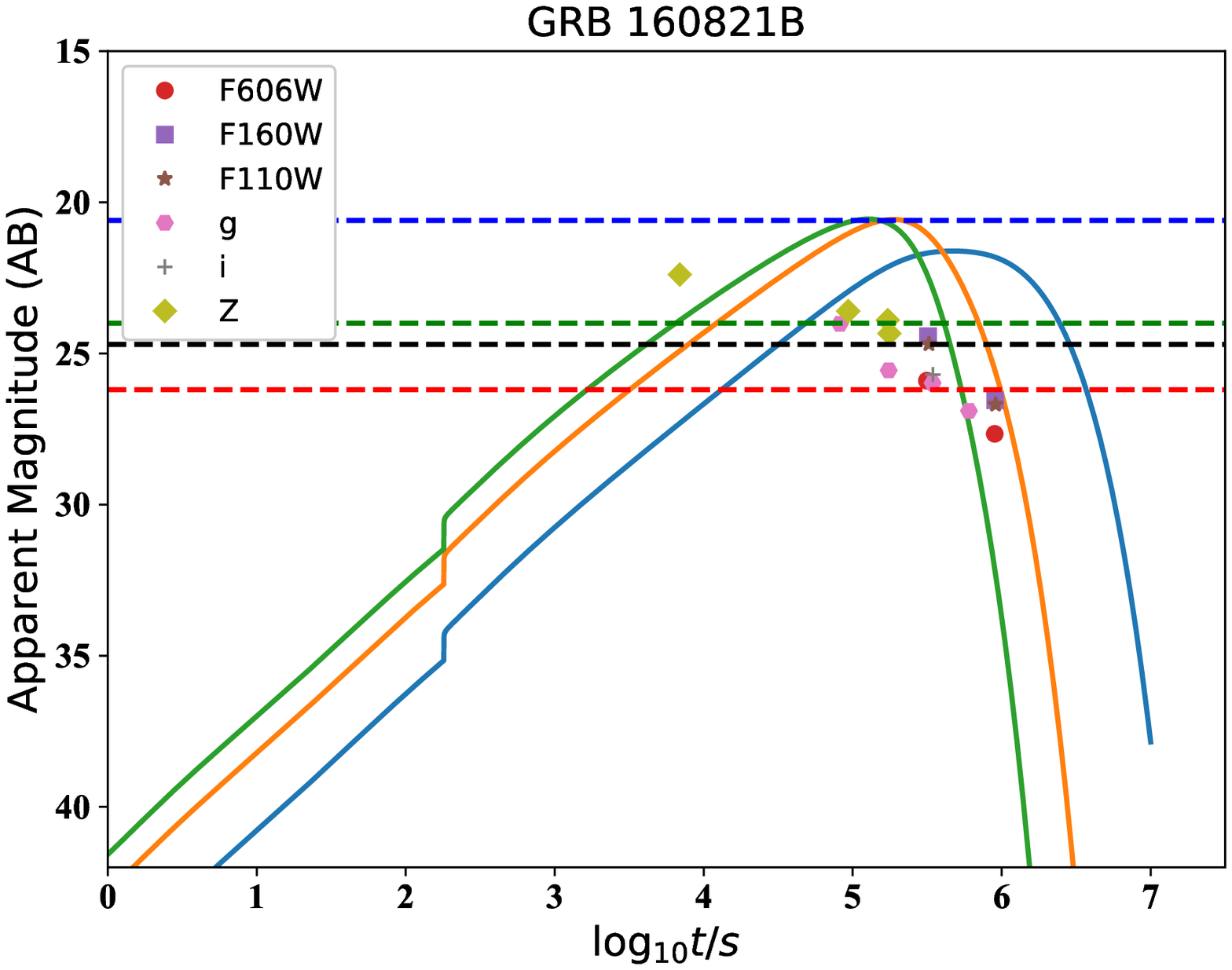}
\includegraphics[width=.25\textwidth]{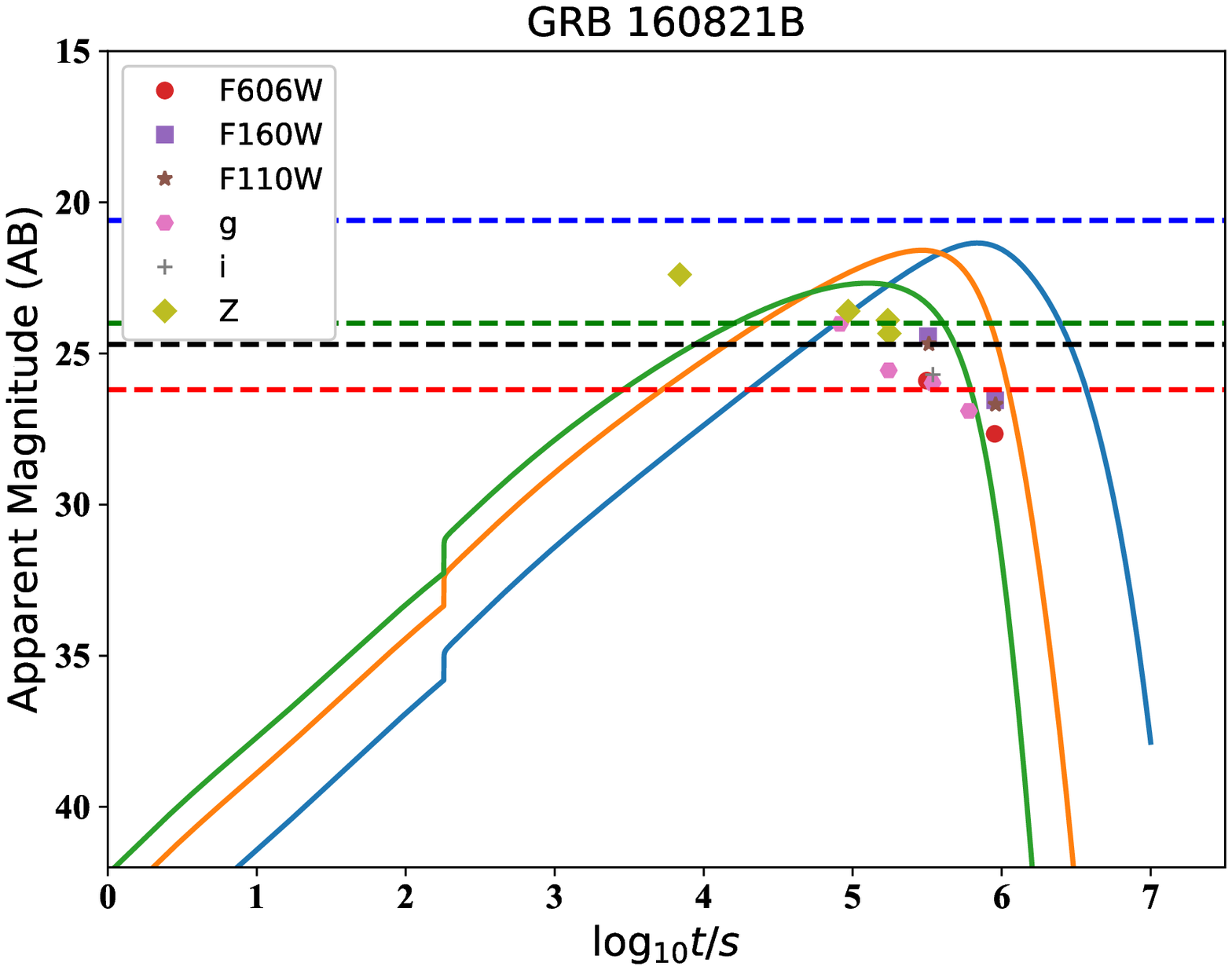}
\caption{Top, middle, and bottom panels are similar with Figure \ref{GRB060801}, but for GRB
160821B. The observed optical data are
taken from Lamb et al. (2019).}
\label{GRB160821B}
\end{figure}


\end{document}